\documentclass[12pt]{article} 

\usepackage{epsfig}
\begin{document} 
%
%\begin{center}

\title{{\bf{Ground state Pseudo-scalar Meson masses\\ in
the Separation Geometry model.}}
\author{G.R. Filewood\\
Research Centre for High Energy Physics,\\
School of Physics,
University of Melbourne,\\
Parkville, Victoria 3052 Australia.}
\date{\today}
\abstract
In this paper the techniques of discretised QCD 
developed in physics/0109024 
are extended to a study of ground-state
pseudo-scalar
mesons  involving higher generation quarks. The correspondence
with empirical values is good with the exception of the
charmed-bottom meson with the predicted value of 6.911GeV
varying considerably from the empirical value of
$6.4\pm0.39\pm0.13$  although
it is still within two sigma and provides an avenue
for further testing of the theory.\\
$\;$\\
$\;\;\;\;$PACS 14.4, 14.80B, 14.80C, 13.40D, 11.30E, 11.30P}
\maketitle

\section{Introduction}

In a previous paper a calculation methodology
was developed to perform discretised QCD mass
calculations. Calculation of the nucleon masses
was performed as an example of the technique.

The purpose of this paper is to extend the calculation
technique developed in the previous paper
to cover quarks of generation higher than the
first.  In this paper we will
examine a selection of $J=0$ ground state
mesons. To do so requires 
additional hypotheses
above and beyond those presented in the
previous paper; the validity of which once again
depends upon the empirical success (or failure!)
of the methodology. We begin by re-examining some
of the basic geometric concepts from the 
previous paper in an abbreviated form.

\section{Basic ideas.}
Consider the symmetry of a coloured cube;

\vspace{0.5cm}

{\centerline
{\epsfig{file=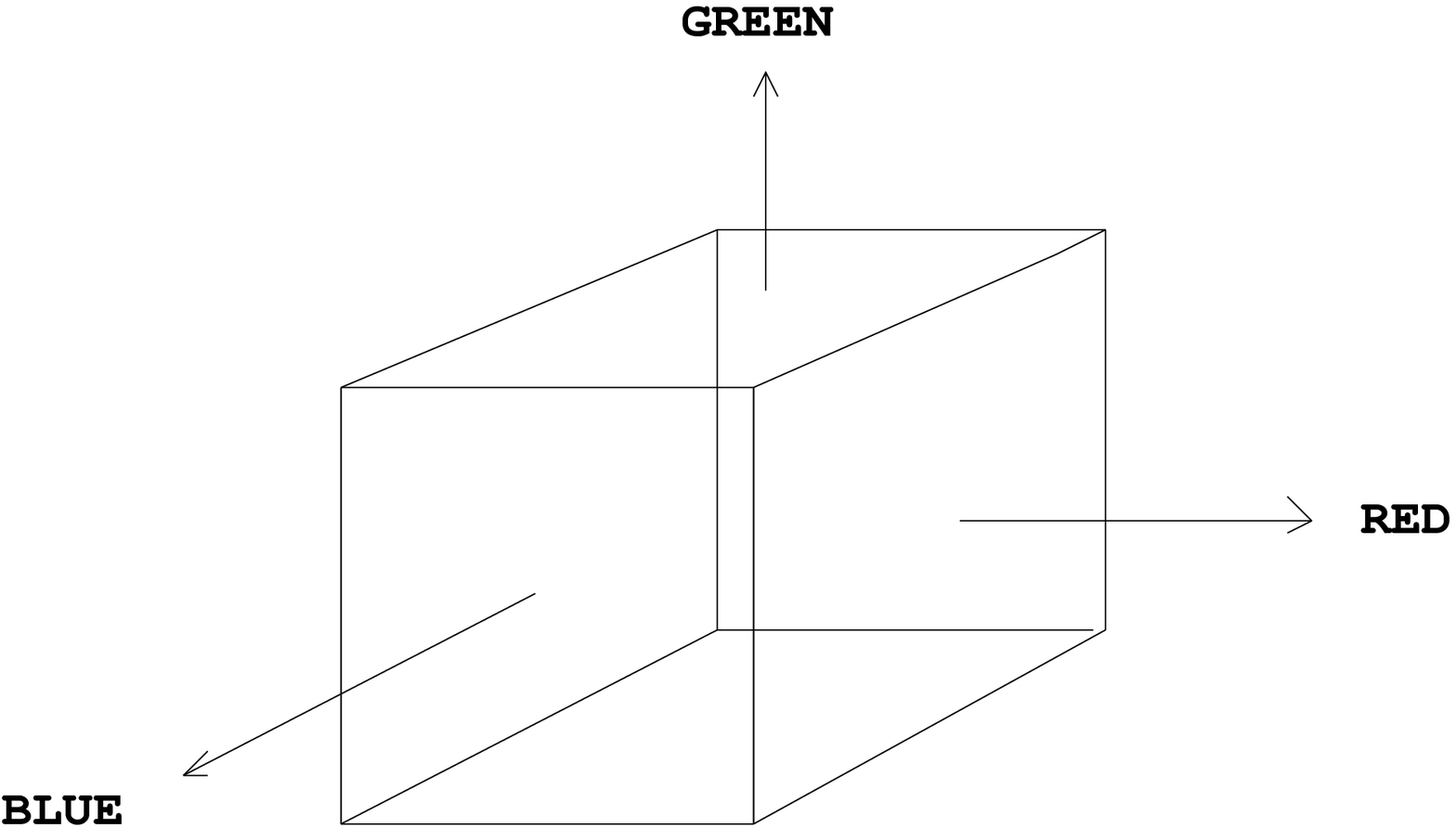,width=9cm}}}

\vspace{0.5cm}

\noindent
where the cube has been given some colour; two 
opposite faces red, two green and two blue. In this
example the observed colour of the cube is a 
function of the position of the cube
(or equivalently of the observer!); in the 
language of contemporary physics it is `gauge
dependent' (although `phase dependent' would, of course, be 
a better description). Indeed, if we imagined that the
rotational symmetries of the cube manifested  as continuous
rotations in the embedding space 
then the observed colour would indeed be
a function of the local phase of the generators of the 
rotations. Alternatively we could consider that the
colour axes are somehow fixed in an `internal'
 space defined by geometry generators and that as the cube 
rotates there is a continuous change in the spectrum  of
colours seen by the observer. The colour itself would 
then be described by a continuous local phase symmetry.

This is the basic idea behind discretised QCD;
that quark `colour' may be represented by
a special kind of `gauged' cubic symmetry. 
In standard theory the `colour' of a quark is not an observable. 
The `colour' of a quark is like phase in electro-magnetic theory; 
local gauge invariance means that the colour is not an observable 
just as the local phase of an electron wave-function is not an 
observable.

In standard QED a phase factor that varies locally (i.e. varies with
spatial position) is not invariant in the absence of a gauge field
because;

\begin{equation}
\partial_{(x)}({e^{i\theta_{(x)}}}\psi)
=i\partial{\theta_{(x)}}{e^{i\theta_{(x)}}}\psi+
{e^{i\theta_{(x)}}}\partial{\psi}
\end{equation}

\noindent
and so, because the first derivative of 
the field enters into the 
Lagrangian, the Lagrangian will not be 
invariant under a local phase 
transformation. In QED if we add the 
photon gauge field, however, 
local phase invariance of the Lagrangian is restored.

In the case of quarks the local phase 
information is more complex
than a simple $U(1)$  $e^{i\theta_{(x)}}$ and 
is governed by $SU(3)_c$ 
symmetry but the principle is the same. However, 
 the gauge particles of QCD, the gluons, 
unlike the photon of QED, carry the charge 
of the underlying field; 
the gluon carry `colour' charge. Moreover, 
instead of the single photon 
the gluons come in eight `colour' varieties. 
There difference are a 
consequence of the non-abelian nature of 
$SU(3)$ and the fact that 
there 
are eight generators  instead of a single $U(1)$ generator.

In discretised QCD the 
`local phase' 
information for 
$SU(3)$ is
carried by a three-dimensional `cube' 
embedded in a four-dimensional
space-time; which is to say that 
the time dimension is 
{\it{undefinable}}
inside the boundary of the cube. 
The discrete phase of the cube -
the spatial position of it's faces, 
edges and vertices, and the length
of the edges, area of square faces 
and three-dimensional volume of the
cube - represents 
colour (and other local gauge) 
information which is rendered 
{\it{unobservable}} by the gauge 
fields; the geometry of the colour 
gauge fields is postulated to be 
in the form of two-dimensional 
`flat-square' geometries analogous to
the square faces of the cube. 
Thus the geometry of a 
gluon in this schema
is that of a `flat square' and it 
contains colour information 
relative to
the surface of the cube from which 
it is emitted and the surface to
which it is absorbed in the 
time-containing embedding space. 
There are 
fundamental theorems in the 
theory which tell us that any 
two-dimensional
geometry will always have zero 
rest-mass so that the `flat-square' 
geometries will always propagate at 
the speed of light. Conversely, 
a three-dimensional geometry such a  
`cubic' quark will always have 
non-zero rest mass.

 With local gauge invariance no geometric
feature of the `cube' structure itself 
will be  directly observable as such - we might consider it
a snapshot `image' of a quark at an instant in time
but to observe it over an interval of time
t=0 requires, by the time-energy uncertainty
relation, an infinite energy probe which of course
is not possible. Can we {\it{indirectly}} infer
the existence of such an underlying symmetry for
quark structure? 

Well, we might try and see if some of the physical
properties
of quarks emerge from a study of such discrete symmetry.
The property which seems to correlate is particle mass
and in a previous paper {\cite{filewood}} this idea
was explored. Discrete subgroups of SU(3)
can be extracted which can be mapped onto the
cubic symmetry - rather like a collection of
discrete `snapshot' images of a continuously
rotating cube and it is found that the  order
of these groups (i.e. the number of elements
in the group) are related to the rest mass.
These discrete sub-groups have  the same basic
symmetry properties of the parent group but in
a discrete form -  so, for example, we might have
three fixed discrete colours for our cube rather
than a continuous spectrum as the generators of the
geometry act.
In the previous  paper the issue of how an
extended object such as a `cube' may be a representation
for an object, such as a quark, which empirically 
behaves like a point was discussed so that issue
will not be revisited here. A brief summary 
of the ideas from {\cite{filewood}} is as follows.

Material particles such as quarks and leptons
are assumed to be represented by a special kind
of geometry termed `affine-set' geometry. The nature of
affine-sets in this schema is related to a decomposition
of the structure of the continuum of space-time. Local
phase factors - U(1), S.U.(2) or S.U.(3) - are assumed 
to be due to this geometry and local gauge invariance
is assumed to be the mechanism whereby this local
phase is rendered unobservable. Geometrically this
becomes equivalent to making the underlying decomposition
of the continuum -
which is present on the {\it{boundary}} of
the geometry -
unobservable. The gauge fields themselves also
relate to the geometry of the boundary.

The mass of any fundamental object is then
a relic of an otherwise empirically unobservable
geometric foundation to physical structure. The fundamental
symmetry of an affine-set geometry is a permutation symmetry
and it is the number of matrix elements representing this
symmetry (always discrete) - minus any massless generators
and modulo any radiative corrections from the gauge fields -
which in general defines the mass; although the situation
becomes quite complex with hadrons.

Conventionally to calculate a hadron mass, such as a 
spin-zero meson, one would postulate a form for 
the potential, insert assumed current quark masses
and then try and determine the (quantum) ground state
energy to define the mass. 
Now whilst this approach may be  completely valid,
and may be necessary for calculating excited states,
in discretised QCD we assume that 
 the discrete symmetry dictates the form and
value that the potential and current quark masses can
take in the ground state by dictating the total energy,
current quark masses and all the symmetry features
that the potential must satisfy.
 To calculate the mass one then determines the 
appropriate components of energy from the symmetry
and then sums them to obtain a mass; which is a  much
simpler approach than working from an assumed
potential  - particularly if the current quark mass
is significantly different to the constituent mass
implying that the quarks 
in the hadron are very relativistic
even in the ground state. Constituent and current quark mass
components are readily calculated in the discretised QCD
approach without the requirement for knowledge about
the velocity of the quarks or  relativistic
corrections.
 The energy of the quarks is simply dictated
by the symmetry and can be written down from the
matrix representation.

The calculation of the mass of a ground state meson
is performed by summing the mass/energy of several
different component parts. These parts
are;

1; constituent quark energy -
this is equivalent to the kinetic energy of
the quarks; always taken `on-shell' at $\alpha_s=1$
(`on-shell' here means a particular energy scale,
called the $T_r$ scale, which is dictated by the 
symmetry),

2; constituent gluon energy - due to the
energy of the gluons exchanged by the quarks,

3; current quark mass -which is the self-energy
of individual quarks and consists of the following
three sub-components;

\hspace{1cm}a; energy equivalence of 
current quark operator content,

\hspace{1cm}b; Higgs scalar components and 

\hspace{1cm}c; potential energy of separation of current quarks.

Each of these parts must be separately calculated and, after
application of radiative corrections, summed to obtain
a final mass. We begin by revising the matrix representations
for the different components.

\section{Matrix representations 1; the $T_r$ group}

$T_r$ is the unique 
 sub-group of SU(2) with 24 elements and
is the S.U.(2) analogue of the tetrahedral
point symmetry (spinor) three-dimensional
group $T_d$. $T_r$ has the following fundamental
representation;

\begin{equation}
\begin{array}{ccccc}
&1&2&3&4\\\\
\alpha:&
\left(
\begin{array}{cc}
1&0\\0&1
\end{array}
\right)&
\left(
\begin{array}{cc}
-1&0\\0&-1
\end{array}
\right)&
\left(
\begin{array}{cc}
0&i\\i&0
\end{array}
\right)&
\left(
\begin{array}{cc}
0&-i\\-i&0
\end{array}
\right)\\\\
\beta:&
\left(
\begin{array}{cc}
i&0\\0&-i
\end{array}
\right)&
\left(
\begin{array}{cc}
-i&0\\0&i
\end{array}
\right)&
\left(
\begin{array}{cc}
0&-1\\1&0
\end{array}
\right)&
\left(
\begin{array}{cc}
0&1\\-1&0
\end{array}
\right)\\\\
\gamma:&
\left(
\begin{array}{cc}
ia&-b\\b&-ia
\end{array}
\right)&
\left(
\begin{array}{cc}
-ia&b\\-b&ia
\end{array}
\right)&
\left(
\begin{array}{cc}
-ia&-b\\b&ia
\end{array}
\right)&
\left(
\begin{array}{cc}
ia&b\\-b&-ia
\end{array}
\right)\\\\
\delta:&
\left(
\begin{array}{cc}
b&ia\\ia&b
\end{array}
\right)&
\left(
\begin{array}{cc}
-b&-ia\\-ia&-b
\end{array}
\right)&
\left(
\begin{array}{cc}
b&-ia\\-ia&b
\end{array}
\right)&
\left(
\begin{array}{cc}
-b&ia\\ia&-b
\end{array}
\right)
\\\\
\epsilon:&
\left(
\begin{array}{cc}
a&ib\\ib&a
\end{array}
\right)&
\left(
\begin{array}{cc}
-a&-ib\\-ib&-a
\end{array}
\right)&
\left(
\begin{array}{cc}
-a&ib\\ib&-a
\end{array}
\right)&
\left(
\begin{array}{cc}
a&-ib\\-ib&a
\end{array}
\right)
\\\\
\phi:&
\left(
\begin{array}{cc}
ib&a\\-a&-ib
\end{array}
\right)&
\left(
\begin{array}{cc}
-ib&-a\\a&ib
\end{array}
\right)&
\left(
\begin{array}{cc}
ib&-a\\a&-ib
\end{array}
\right)&
\left(
\begin{array}{cc}
-ib&a\\-a&ib
\end{array}
\right)
\end{array}
\end{equation}

\noindent
where $a={1\over2}$ and $b={{\sqrt3}\over2}$.
The elements of these matrices are built up
from the roots of unity; square roots, fourth 
roots and sixth roots. These elements close to
form a group with 24 elements which we shall
designate $T_r$. The generators of the
group may be taken as the $\gamma_4$ and
$\delta_4$ matrices. Note that, although it is a
sub-group of 
 the continuous version $SU(2)$, $T_r$ has only
two, instead of three, generators. 

$T_r$ group is a kind of discrete spinor group.
It is analogous to the geometric group
$T_d$ familiar to physical chemists which is the
group of the tetrahedron in three-dimensions with
24 elements and 2 generators.($T_d$ is a point space
group in three dimensions 
familiar to physical chemists which is spinorial
but is not isomorphic to $T_r$).

The {\it{mass-equivalence}}
 of $T_r$ is defined as ${\cal{R}}.(4!-2)$
where (4!-2) is the number of {\it{non-generator}}  elements
in the group and $\cal{R}$ incorporates any radiative
corrections due to the elevation of the cardinality
of the boundary of the geometry to the continuum
by the gauge fields. In the transition to
a  field theory analogue the
discrete pair of $T_r$ generators 
are assumed to   manifest as  continuous
 (massless) intrinsic spin generators and this is 
the source of the $-2$ in the above definition; 4!
is of course simply the order of the $T_r$ group.

\section{Matrix representations 2; embedding $T_r$ in SU(3)}

Quarks are assumed to have a cubic symmetry in 
affine-set geometry in discretised QCD. Just as 
with the analogy between $T_d$ and $T_r$ we wish
to find a way of embedding a point-symmetry
representation of a cube in a complex space.

We can parallel the root structure of SU(3) by
embedding three copies of  SU(2) 
in SU(3). We can  embed three copies of the 
$T_r\;\{\lambda_{ij}\}$
group as follows to form a discrete sub-group
of SU(3);
\begin{equation}r=\left(
\begin{array}{ccc}
1&0&0\\
0&\lambda_{11}&\lambda_{12}\\
0&\lambda_{21}&\lambda_{22}
\end{array}
\right)
\;;\;g=
\left(
\begin{array}{ccc}
\lambda_{11}&0&\lambda_{12}\\
0&1&0\\
\lambda_{21}&0&\lambda_{22}
\end{array}\right)\;;\;
b=
\left(
\begin{array}{ccc}
\lambda_{11}&\lambda_{12}&0\\
\lambda_{21}&\lambda_{22}&0\\
0&0&1
\end{array}
\right)
\end{equation}

\noindent
and we also form the following `colour-dual'
embeddings;

\begin{equation}\bar{r}=\left(
\begin{array}{ccc}
-1&0&0\\
0&i\lambda_{11}&i\lambda_{12}\\
0&i\lambda_{21}&i\lambda_{22}
\end{array}
\right)
\;;\;\bar{g}=
\left(
\begin{array}{ccc}
i\lambda_{11}&0&i\lambda_{12}\\
0&-1&0\\
i\lambda_{21}&0&i\lambda_{22}
\end{array}\right)\;;\;
\bar{b}=
\left(
\begin{array}{ccc}
i\lambda_{11}&i\lambda_{12}&0\\
i\lambda_{21}&i\lambda_{22}&0\\
0&0&-1
\end{array}
\right)
\end{equation}

\noindent
The three `colour' matrices
$r,g\;\mbox{and}\;b$ plus the `colour-dual'
basis $\bar{r},\bar{g}\;\mbox{and}\;\bar{b}$
together form a discrete version of the 
 group of $SU(3)_c$.
Also notice that each set of colour matrices
forms a group with 24 elements. The `dual' or
`colour-bar' matrices alone do not form a group
since the product of any two colour-bar
matrices is a colour matrix. However, the
combination of $c_i+\bar{c}_i$ (for colour index
i) forms a group
with 48 elements and two generators located
in the colour-bar set of elements. 
Let us
call this group $T_c$. It is also a 
`discrete spinor' 
 analogous to the $T_r$ group;
\begin{equation}
(i\gamma_4.i\delta_4)^2=-I_2
\end{equation}
Note that the colour-dual is not the same
as the Hermitian conjugate matrix.

We might propose $T_c$  as a model
for a Skyrmion field of charge 4; i.e.
one with cubic symmetry. 
A corresponding point symmetry
(spinor) discrete group would be $O_h$
which has 48 elements and two generators
but, again like the $T_d\;{\rightarrow}T_r$
analogy, is not isomorphic to $T_c$. We will see 
later how the geometric analogy works
but to obtain a picture think
of the three colours as represented
by the three orthogonal pairs of opposite faces of
the cube; one pair of faces representing each
of the three possible  $T_{c_i}$ 
groups with the four vertices of
each of the two square faces 
representing each tetrahedral
sub-group. To represent the total geometry
we form a `particle vector' which is composed
of three components. First form a matrix
representation of $T_{c_i}$ as follows;
\begin{equation}
R=\left(\begin{array}{cc}
r&0\\0&\bar{r}\end{array}
\right),
G=\left(\begin{array}{cc}
g&0\\0&\bar{g}\end{array}
\right),
B=\left(\begin{array}{cc}
b&0\\0&\bar{b}\end{array}
\right)
\end{equation}
\noindent
and then form the three-component particle
vector $\stackrel{\rightarrow}{P}=
(R,G,B)$.
(Of course the order of the components here
is arbitrary).

Now define the following `current-quark' operators;
\begin{equation}
q_r=\left(
\begin{array}{ccc}
-1&0&0\\
0&i&0\\0&0&i\end{array}
\right),
q_g=\left(
\begin{array}{ccc}
i&0&0\\
0&-1&0\\0&0&i\end{array}
\right),
q_b=\left(
\begin{array}{ccc}
i&0&0\\
0&i&0\\0&0&-1\end{array}
\right)
\end{equation}
\noindent

Lastly we require what is referred to in the
text as a `script identity' $\cal{I}$. It carries
a colour index;
\begin{equation}
{\cal{I}}_r=\left(
\begin{array}{ccc}
1&0&0\\
0&-1&0\\
0&0&-1\end{array}\right),
{\cal{I}}_g=\left(
\begin{array}{ccc}
-1&0&0\\
0&1&0\\
0&0&-1\end{array}\right),
{\cal{I}}_b=\left(
\begin{array}{ccc}
-1&0&0\\
0&-1&0\\
0&0&1\end{array}\right)
\end{equation}

The purpose of these constructions will become
apparent as follows. We form operators from the
$q_i$'s and $\cal{I}$'s as follows (the following
examples are red format but the other colours follow
suit);
\begin{equation}
\begin{array}{cc}
U_r&D_r\\\\
\left(
\begin{array}{cc}
{\cal{I}}_r&0\\
0&I_3\end{array}\right)&
\left(
\begin{array}{cc}
q^*_r&0\\
0&q_r\end{array}\right)\\\\
\left(
\begin{array}{cc}
q^*_g&0\\
0&q^*_g\end{array}\right)&
\left(
\begin{array}{cc}
I_3&0\\
0&I_3\end{array}\right)
\\\\
\left(
\begin{array}{cc}
q^*_b&0\\
0&q^*_b\end{array}
\right)
&
\left(
\begin{array}{cc}
I_3&0\\
0&I_3\end{array}
\right)
\end{array}
\end{equation}

\noindent
where each operator has three
components represented above in column format.

Two things need mentioning at this stage;
1; the quark operators $q_i$ and ${\cal{I}}_i$
are assumed confined to their respective colour
space i; that is, the $q_r$ operators only operate
on the red particle vector  matrix etc, 
and 2; the order of the 
application of the operators is of no significance
because they are diagonal matrices.

These `operators' will represent the current quark content
of a hadron and the `particle vector' will determine
the constituent mass/energy.

\section{Matrix representations 3; glue}

Discretised gluon operators are represented
as follows;
\begin{equation}
b\;\left(\begin{array}{cc}
I_3&0\\
0&q_bq^*_g
\end{array}
\right)
g^*.
\left(
\begin{array}{cc}
g&0\\0&\bar{g}
\end{array}
\right)
=
\left(
\begin{array}{cc}
b&0\\0&\bar{b}
\end{array}
\right)
\end{equation}

\noindent
where * represents the complex conjugate.
Similar upper product 
operators are defined for example;

\begin{equation}
r\;\left(\begin{array}{cc}
q_rq_g^*&0\\
0&I_3
\end{array}
\right)
g^*.
\left(
\begin{array}{cc}
\bar{g}&0\\0&{g}
\end{array}
\right)
=
\left(
\begin{array}{cc}
\bar{r}&0\\0&{r}
\end{array}
\right)
\end{equation}

These are then combined to form 
operators which operate on the particle
vector $\stackrel{\rightarrow}{P}$
for example;
\begin{equation}
\left(
\begin{array}{c}
g\;\left(\begin{array}{cc}
I_3&0\\
0&q_gq^*_r
\end{array}\right)
r^*.\\\\
r\;\left(\begin{array}{cc}
I_3&0\\
0&q_rq^*_g
\end{array}
\right)
g^*.\\\\
\left(\begin{array}{cc}
I_3&0\\0&I_3
\end{array}\right)
\end{array}
\right)
\end{equation}

\noindent
when applied to the particle vector
$\stackrel{\rightarrow}{P}=(R,G,B)$
produces $(G,R,B)$ i.e. this is a R-G
gluon.

\section{Matrix representations 4; current Quarks.}

 The current quarks are built
up from the quark operators used previously
to define the particle types. The current quark
masses are the corresponding masses of these
operators plus an extra piece whose origin may be 
the Higgs field but which can be interpreted in
terms of a potential of separation  between current quarks.

We can split up the operators into two
pieces; which is convenient for calculations.
Each current quark operator consists of a vector composed
of 3 blocks $6\times6$ matrices. Each $6\times6$
block is composed of two $3{\times}3$ blocks on
the diagonal. We 
separate out
the upper $3{\times}3$ blocks as one set of
matrices and the lower $3{\times}3$ set of 
blocks as another and represent them as
follows (the representation below is for
the proton as the reader can confirm
by applying the operators to the 3-quark (first generation
only)column
particle vector $\stackrel{\rightarrow}{P}=
(R,G,B)$) remembering that the red operators
operate only on the red $\stackrel{\rightarrow}{P}$
components etc.;

\vspace{0.5cm}

\begin{equation}
\mbox{strong component}=
\begin {array}{c}
red\rightarrow\\
green\rightarrow\\
blue\rightarrow
\end{array}
\left(
\begin{array}{ccc}
\cal{I}&q^*&I\\
q^*&\cal{I}&I\\
q^*&q^*&q^*\\
\end{array}
\right),
\;\;\;
\mbox{E.M. component}=
\left(
\begin{array}{ccc}
{I}&q^*&I\\
q^*&{I}&I\\
q^*&q^*&q\\
\end{array}
\right)
\end{equation}

\vspace{0.5cm}

The rows of these matrices are the three
colours (order red, green and blue in the
notation used in this paper from top down)
and the columns are the type of quark;
the left hand and central column of each
matrix above is an up-type quark and the right
hand column is a down-type quark. The up-type
quarks have their colour defined by the colour
of the single identity they carry. Thus for example
the first column of these matrices represents an 
 up red  quark. The single q or q* in the down defines its
colour so that the last column above is a blue down
quark in each case.

The strong components couples to the $C_i$ colour matrix
of the particle vector which does not contain
any generators. The E.M. component couples
to the $\bar{C}_i$ matrix with two generators.
(This coupling is required to properly define the
charge characteristic of the particle in the given 
representation).
The order of a $C$ matrix is 24 elements. The
order of a $\bar{C}$ matrix is (4!-2) elements
where, as in the case of the $T_r$ group, the
generators are assumed to be massless intrinsic
spin generators in the transition to a continuum
field theory. 

The masses
of the current quark operators are  defined in
terms of the matrices to which they couple.
The exception to the rule applies to the
identities in the E.M. component which are
spinless and of order 4! (i.e. the generators
of $\bar{C}$ are massive under the identity
components of the current quarks - the I
or the $\cal{I}$ components). The identities
in the strong component act like `strong charges'
and carry a relative mass sign. For example, in
the proton mass listed the two identities cancel
the two script identities leaving only the net 
5 q*'s to contribute to the `strong' matrix  mass.

\section{The Higgs sector}

This is the least satisfactory part of
the theory but some progress can be
made with a combination of a few 
simple suppositions and comparison with
empirical data. There are some interesting
hints of new physics in the discretised model
of the Higgs sector and in what follows we
shall mention these in passing.  We divert initially to study
the vector gauge boson masses because these
provide some support for what follows with
the fermions.

There seems to be a general rule in affine-set
geometry relating the the {\it{maximal massless 
subgroup}} of a geometry to possible unitary symmetries
in the transition to a field theory
although we have no formal mathematical
proof of the connection. Let us
look at this in terms of the $T_r$ 
tetrahedral group. This is
the group which in the previous paper 
\cite{filewood} was used to
underpin lepton (and indirectly viz the $T_c$
analogue also quark) structure.
The  permutation group 
corresponding to $T_r$  is $S_4$ 
- the characteristic symmetry of a
four-point   affine-set geometry.

If we form a mental picture of a tetrahedron
we note that it has four 3-vertex subsets
forming four triangular faces. To see that there is an
orbit of the $T_r$ group with three elements note
that; 
\[
(\epsilon_2)^3=I_2
\]
where $\epsilon_2$ is one specific matrix  listed in the
table of $T_r$ group matrices.
Here $\epsilon_2$ is the analogue of a $C_3$ rotation in
a three-dimensional tetrahedron described by
the group $T_d$ familiar to physical chemists. The
continuous analogue of this will be a U(1) rotation
in a two-dimensional plane.
The stabiliser of this orbit is the single remaining
point of the set of four defining the geometry
and this sets the orientation of the associated 
U(1) symmetry axis. The affine-set
tetrahedron in the previous paper 
was associated with charged lepton masses
and the three-point sub-group associated with
the photon geometry. That is, the photon geometry
is the maximal massless subgroup of the $S_4$
tetrahedral group and the single stabiliser point
dictates the associated symmetry as U(1).

Without proof we may seek to generalise this result
to scalar geometries. Scalar affine-set geometries have
the peculiar property that any sub-set of points 
always defines a permutation group. There will always be
a maximal massless set which will form an orbit
with the remaining points forming the associated 
stabiliser. We postulate that this stabiliser, when part
of a scalar set of points, will always define a unitary
symmetry as a function of the number of points in the
stabiliser. We express this more precisely as follows;

\vspace{0.5cm}

{\underline{Hypothesis}; if a scalar affine-set geometry 
of order $S_n$ exists then its' corresponding
unitary symmetry group is at most SU(n-m) where 
the index
(n-m) is the number
of points of the affine-set geometry present in the stabiliser 
of the orbit
of the maximal massless sub-group containing m points  and
(n-m) is  a prime number; or,
if (n-m) is not a prime number, then 
the geometry defines  a product
of unitary groups of prime index whose sum of indices is
(n-m).

\vspace{0.5cm}

For $S_8$ (fermionic!)cubic symmetry
 it is interesting to note that,
although we have previously seen that it is
a spinor geometry not a scalar,
in the case of the strong interaction
in colour space the
gluon geometry is a `flat' (two-dimensional and
hence massless; see {\cite{filewood}
for more detail}) 4-point geometry which can represent
a single square face of the cube. 
In colour space this is the maximal massless
subgroup of degree $S_4$. Hence m=4, n=8
and the  decomposition is 
either to SU(2)xSU(2) or SU(3)xU(1); the latter of
these two prevails empirically -
presumably because the U(1) subgroup
is specified by the requirement that
the quarks are electro-magnetically charged.

Now, for the standard model we have SU(3)xSU(2)xU(1)
so that m=6. What is the appropriate value of n?
In affine-set theory all massive fermions decompose 
to $T_r$  ($S_4$) equivalent units (for example, in calculating
the mass of a cubic quark in previous sections
the cube was decomposed into $T_r$ units which were then summed).
 In the absence of the
Higgs field, standard model fermions are massless
so the obvious choice is to take n-m=4;
that is, setting the fermions as massless and
representing the maximal massless subgroup - so that
n=10 or an $S_{10}$ scalar affine-set geometry
with 10! elements represents the standard
model prior to spontaneous symmetry breaking. Now,
when we give the fermions mass there will be
a shift in the maximal massless sub-group from
n-m=4 to n-m=3 where the triangle $S_3$ symmetry
is the photon group. (Note that 
the persistence of a massless
object of order $S_4$
- gluons -  is unique to colour space
and not a $T_r$ (spinor)  representation; the issue
of the masses of neutrinos is more complex -
in the case at hand we are only concerned with the
usual electro-weak symmetry breaking problem of 
ensuring that the photon is massless but the standard
model charged fermions
gain mass).
This shift in maximal massless sub-group may be the
affine-set theory analogue of spontaneous symmetry
breaking and can be analysed in discrete theory
in the following way.

A factorial number has a  decomposition
into factorial sub-groups;

\begin{equation}
n!=\sum_{k=1}^{n-1}k.k!+0!
\label{decomp}
\end{equation}

Symmetry breaking will demand a
decomposition of the $S_{10}$ symmetry to
sub-groups of order no more than $S_9$.
(This is because the standard model has 
n=6 and for the photon m=3). Now, a scalar 
affine set can be decomposed in the pattern of
(\ref{decomp}) because, in the absence of any 
spin generators, every subset of points defines
a subgroup which is always a permutation group.
The
symmetry breaking can be effected by discarding
the massless pieces of order $S_3$ (or less),
which gives a {\it{minimal}} decomposition
into exactly
 six massive components
(remember we have discarded the
massless pieces);

\begin{equation}
S_{10}\rightarrow
9.S_9+8.S_8+7.S_7+6.S_6+5.S_5+4.S_4
\label{Hig}
\end{equation}

The photon is now the maximal massless geometry
of order $S_3$ and the possible unitary symmetries 
by hypothesis 9 are (neglecting the possibility of
multiple U(1) subgroups and allowing only groups of
prime index);
\[
\begin{array}{cc}
A:&
S_9{\rightarrow}SU(3){\times}SU(3)\;\mbox{or}\;
SU(3){\times}SU(2){\times}U(1)\;\mbox{or}\;
SU(2){\times}SU(2){\times}SU(2)
\\
B:&
S_8{\rightarrow}SU(5)
\\
C:&
S_7{\rightarrow}SU(2){\times}SU(2)\;\mbox{or}\;
SU(3){\times}U(1)
\\
D:&
S_6{\rightarrow}SU(3)\\
E:&
S_5{\rightarrow}SU(2)\\
F:&S_4{\rightarrow}U(1)
\end{array}
\]

It appears that a decomposition into a product of unitary 
groups requires a {\it{gauged}} symmetry and that the scalar
Higgs components are not gaugable as such in their couplings
to the fermions. The
{\it{physical}} explanation for this is that the scalar
Higgs components for the fermions `live' in the interior
space of the geometry (e.g. `inside' the cube for a quark)
and that this space does not contain the time dimension and
cannot be gauged for this reason. Comparison
with empirical data shows that,
for the Higgs field - fermion mass couplings,
 possibilities A and C do not occur whilst
B,D,E and F appear to manifest physically.
(Possibility A seems to contains the standard model decomposition
for the non-Higgs gauged sector but this was by construction;
see above). It is particularly
interesting that, as we shall see, the empirical correlation with
 possibility B is good which may be the 
first empirical suggestion of the existence of an $SU(5)$ symmetry
in nature.

The situation is complicated because of the requirement
for a  propagating Higgs field to render the electro-weak
theory unitary. This requirement is specifically in relation
to the vector gauge bosons which require the Higgs field
in their radiative corrections to preserve unitarity. This
is not possible if the Higgs components are `buried outside time'
as it were in the `interior' timeless space of three-dimensional
affine geometries representing the massive vector gauge
bosons! Indeed, as we shall see, comparison with
empirical data does indeed show that the Higgs propagates
in relation to the vector gauge bosons which we will now study in
some detail.

 We know that, for completeness and consistency,
we require a model for the massive vector gauge 
bosons. $T_r$ is a spinor group and unsuitable
but there is a three-dimensional point symmetry 
group $T$ which has only vector generators.
$T$ has 10 non-generator elements and 2 generators
but 12 elements do not represent an $S_n$ symmetry
of an affine set ($S_3$ for example has 6 elements
whilst $S_4$ has 24 elements). However, if we 
discard the massless photon orbit from the
scalar $S_{10}$ the $S_4$ scalar subgroup which results 
will acquire mass
through a decomposition of the Higgs field. This 
decomposition is an inevitable consequence of affine
sets.This could be 
used to couple to a four-point {\it{bosonic}} $T$
geometry to give it 
an affine symmetry (i.e. re-define its'  elements
as a permutation group).
Thus we might then try to couple this scalar quantity to
a $T$ geometry to define it by chopping it
in half to mirror the fact that $T$ has 12 
instead of the $S_4=4!$ elements of the affine
set of a four-point (spinor) geometry. This seems to work.
The other mass
components we must assume couple to the geometry
rather like `lumps of lead' and are not  distinctly
related to the $T$ morphology - although 
comparison with empirical data shows that this
does not apply to the (scalar not fermionic!)
$S_8$ sub-component  which  is
also chopped in half. One may account for this
as relating to a coupling to a `$T$'bosonic-version of
the quark $T_c$ geometry.  Empirically
this gives a very good  fit to the data i.e.
with the singular assumption that
there is mixing, which is maximal,
at the $S_8$ (quark) level  and  also at the 
$S_4$ (lepton) level; which intuitively is
physically sensible. The way we do it is, given
that  the number 
of elements of 
$T$ is half that of $T_r$, we assume  ${1\over2}.4.S_4$ may be 
regarded as defining 4 units of (bosonic) $T$ as distinct 
from 2 units of (fermionic) $T_r$. Similarly, we may postulate
a bosonic  point symmetry analogue of the fermionic
 $T_c$ group with half the number of elements 
in each $T_r$ sub-group analogue (i.e. $C_i$ and $\bar{C}_i$)
and representing
the point group $O$ of the cube with vector generators.
Comparison with empirical data gives 
the mixing  represented  as;

\vspace{0.5cm}
\begin{center}
\[
S_{10}\rightarrow
9.S_9+8.S_8+7.S_7+6.S_6+5.S_5+4.S_4
\]
\begin{tabular}{|c|c|c|c|c|c|c|}\hline
Boson&9.$S_9$&$8.S_8$&$7.S_7$&$6.S_6$&$5.S_5$&$4.S_4$\\\hline
$W^+$&1&1/2&1&1&1&1/2\\\hline
$W^-$&1&1/2&1&1&1&1/2\\\hline
$Z^0$&1&2&1&1&1&2\\\hline
\end{tabular}
\end{center}

\vspace{0.5cm}

\noindent
Here each boson is represented by an $S_{10}$ permutation
symmetry based on a permutation of the ten non-generator
elements of the discrete (bosonic) group $T$. The decomposition
results from rendering the photon massless which pulls out
the $S_3$ subgroups leaving only the massive pieces of order
$S_4$ or greater.
Notice how mass components have shifted from the 
$W^{\pm}$ to the $Z^0$ with 50\% of the charged
boson $S_8$ component shifting across from 
the $W^{\pm}$ to the $Z^0$. The resulting electro-weak mixing
angle can be precisely calculated but it is interesting to note
that, in spite of the value of the angle, it this theory
it results from maximal mixing! 
We proceed with a calculation of the
predicted masses of the bosons from the above table.

The W bosons acquire radiative corrections to
their mass. The standard affine-set geometry
correction is applied; it is non-perturbative 
(see \cite{filewood});
${\cal{R}}=1+\alpha_{(q^2=M_W^2)}$
and the shifted (mixed into the $Z^o$)
components keep their
radiative corrections. Here the Higgs components 
acquire radiative corrections; unlike the case for the 
fermions the Higgs field propagates in space-time
in the case of the mass components of the $W$ bosons
(necessary for unitarity)
and those components mixed from the $W$'s into the $Z$. 
Comparison with physical data shows that the
Higgs components native to the $Z^o$ do not propagate
in the theory and do not acquire radiative corrections! The 
physical implications of this are quite astonishing. (It implies,
for example, that spin statistics and space-time structure
are not independent and that electro-magnetism has
an interpretation in terms of space-time structure - this arises because
of the double whammy implied by the data correspondence;
 to a high degree of precision
the data correspondence with theory shows that the Higgs components
{\it{do not}} propagate with respect to the fermions (they are buried
inside the geometry and outside the time dimension). This is represented
by the lack of radiative correction to the mass components of 
the fermions derived directly from the Higgs sector.
 Only  for the charged fundamental (i.e. no fermionic
sub-structure) bosons does the Higgs 
appear to propagate in space-time;
 but only in respect to the 
fundamental boson  carrying electro-magnetic charge. 
We must speculate that
 the mixed 
component from the W's to the Z are sufficient to confer unitarity 
on the Z. The radical interpretation, which I favour, is that
the Higgs changes the statistics of space-time; that is, it acts as 
a supersymmetry operator on space-time structure (as distinct from 
supersymmetry in the particle spectrum) and generates electro-magnetic
charge in the process. If this is true we would not expect to find
a `free' Higgs floating around in a particle accelerator because
its' existence would violate charge conservation even though it is 
electrically neutral!

Now,
$\alpha_{(q^2=M_W^2)}\approx128^{-1}$
and the calculation proceeds by summing the number of 
matrix elements of the groups in question
(the matrix order will be equal to the number
of group
elements). Calculation 
from the above table gives the W mass as;
\[
W_M^{\pm}=(1+\alpha_{q{\approx}M_W^2}).3467400\]
and the $Z^0$ mass as; 
\[
Z_M^0=3628776+(1+\alpha_{q{\approx}M_W^2}).322656\]
 where the numbers
are obtained by summing over the number of elements in 
the respective groups.
 (For example, $9S_9$ has 9.9!=3265920 elements).
These numbers are converted to MeV by using the 
mass equivalence of the $T_r$ unit which is exactly
equal to the electron rest mass;
$T_r=(1+\alpha_{q^2=m_e^2}+G_f)(4!-2)\approx0.511MeV$
and $G_f$ is a dimensionless number which reflects
the relative strength of the weak interaction at low energy;
we may take 
$(1+\alpha_{q^2=m_e^2}+G_f)\approx1.0073115$.
One then readily obtains;

\[M_W=80.578GeV(80.41(0.10))\]
and
\[
M_Z=91.173GeV(91.187(0.007))\]

and
\[sin\theta_w=\]

This provides empirical support for the claim that 
the breakdown components of the $S_{10}$ symmetry
listed previously do indeed represent the Higgs field.

If a 3 `units' of $S_{10}$ Higgs couples to the vector gauge
bosons which represent the adjoint rep. of
S.U.(2) then we might  expect a 2 units of `Higgs'
to specify the fermion masses - at least this is
the supposition we will make.  As before 
requiring  a vacuum structure with
massless photons (and gravitons
; also ignoring for the moment the existence
of the massless gluon which we must treat as somehow
separate to   spontaneous symmetry breaking)
irreversibly breaks  down the $S_{10}$ scalar
into six components. 
If we start with two $S_{10}$
units this means that in the quark sector 
a maximum of six objects can acquire mass 
and in the lepton sector
another six objects can acquire mass; the six masses in 
each case will have a hierarchy based on eq.({\ref{Hig}}).
 In the situation of
simplest symmetry the tau neutrino has the same mass as
the top-quark  etc. as per the following table;

\vspace{0.5cm}

\begin{center}
\begin{tabular}{|c|c|c|c|c|c|c|}\hline
Fermion&$9.S_9$&$8.S_8$&$7.S_7$&$6.S_6$&$5.S_5$&$4.S_4$\\\hline
$top$&1&0&0&0&0&0\\\hline
$bottom$&0&1&0&0&0&0\\\hline
$charm$&0&0&1&0&0&0\\\hline
$strange$&0&0&0&1&0&0\\\hline
$up$&0&0&0&0&1&0\\\hline
$down$&0&0&0&0&0&1\\\hline
$\nu_{\tau}$&1&0&0&0&0&0\\\hline
$\tau$&0&1&0&0&0&0\\\hline
$\nu_{\mu}$&0&0&1&0&0&0\\\hline
$\mu$&0&0&0&1&0&0\\\hline
$\nu_{e}$&0&0&0&0&1&0\\\hline
$e$&0&0&0&0&0&1\\\hline
\end{tabular}
\end{center}

\vspace{0.5cm}

\noindent
where one `Higgs' unit of mass  has been used to generate the 
quark masses and one to generate the lepton masses
as per eq.(\ref{Hig}). (Of course we would need another
two `Higgs' if we wanted to also give mass to equivalent
sets of anti-particles).
However, the geometry of the neutrinos in the 
theory is `flat' which means they can only have
masses generated radiatively; they cannot acquire 
`volume' masses which are characteristic of the
the Higgs components
(a scalar affine-set geometry of four or
more points will always define at least a 
three-dimensional space. Thus the above table is badly
broken. We may assume that mass components
cannot be shifted `horizontally' between columns because to
do so would require another round of symmetry breaking
to strip-off more massless sub-units.
Clearly all the mass from the listed $\nu_{\tau}$
will appear in the top (it has nowhere else to go).
Using   a number of different inputs
including theoretical and empirical (for example the
known lepton masses have been used to help construct
the following table) it is possible to determine the correct
form of the quark-lepton mass mixing matrix;

\vspace{0.5cm}

\begin{center}
\begin{tabular}{|c|c|c|c|c|c|c|}\hline
Fermion&9.9!&8.8!&7.7!&6.6!&5.5!&4.4!\\\hline
$top$&2&10/8&0&0&0&0\\\hline
$bottom$&0&5/8&0&0&0&0\\\hline
$charm$&0&0&1&3/6&0&0\\\hline
$strange$&0&0&0&1/6&3/5&0\\\hline
$up$&0&0&0&0&0&0\\\hline
$down$&0&0&0&0&0&0\\\hline
$\nu_{\tau}$&0&0&0&0&0&0\\\hline
$\tau$&0&1/8&1&2/6&0&0\\\hline
$\nu_{\mu}$&0&0&0&0&0&0\\\hline
$\mu$&0&0&0&1&2/5&0\\\hline
$\nu_{e}$&0&0&0&0&0&0\\\hline
$e$&0&0&0&0&0&0\\\hline
\end{tabular}
\end{center}

\vspace{0.5cm}

The first thing to notice is that, no sooner had
we argued that components cannot be shifted
`horizontally' than a $5.S_5$ disappears from 
the up/$\nu_e$ column entries.
We will modify the accounting later to
bring the `bboks back into balance'.
 We might have expected
a `1' at the up on-diagonal
entry; and indeed in the charged pions
appear to acquire an $S_5$ mass component
as an SU(2) singlet (see the calculation
of the charged pion mass given in subsequent sections)
but the mixing of the components of the first
generation quarks and leptons is a little complicated
and not well understood but
 we will look at it more closely later. 
For the moment simply regard the above table as
a semi-empirical analysis. 

The Higgs components
are lopsided towards the quark sector precisely because of
the masslessness 
(with respect to the Higgs mechanism)
of the neutrinos. One might have expected
a 9.9! for the $\nu_{\tau}$ but if the neutrinos are massless
(or have masses only generated radiatively)
then this component is picked up by the top quark which has a
2 in the first column. The 8.8! has broken up in the S.U.(5)
multiplets (1+5+10).8!=2.8.8! in descending order of mass from
the top. This decomposition is in keeping
with hypothesis 9 which specifies SU(5)
as the appropriate symmetry for the stabiliser of
the 
$S_8$ terms
under a massless photon. In fact (1+5+10) is the only breaking
possible here if S.U.(5) dictates the allowed components
(there are only three available `slots' if the neutrino
is massless and the sum of terms must
be 16). This symmetry is apparently not a gauged symmetry and
appears quite distinct to familiar local-gauge 
symmetries. The tau gets a singlet S.U.(5) rep of mass
and  there is mixing
of mass components between the top and the bottom based
on the S.U.(5) multiplets 5 and 10. No Higgs mass shifts between
columns but there is effective 
quark mass mixing because the Top
acquires a component from the column diagonal for the bottom.
The list gives the top quark mass as $\approx$ 160GeV. and
the bottom mass as $\approx$ 4.65GeV. These are the dominant,
but not the only, components of the `current-quark'
mass of these quarks.

The $S_7$ symmetry is unbroken. The neutrinos
 have no Higgs components
of mass in this theory and the lepton sector 
7.7! units of elements shifts to the tau; the only available
slot. The third tau term is interpreted as the sum of
two singlets. The other entries follow the appropriate
multiplets of the unitary groups appropriate for 
the corresponding decomposition given by hypothesis 9. 

 Note that there is  {\it{inter}}-generational mixing of mass
components in the lepton sector with the tau extending into the
muon column. This inter-generational mass-mixing,
seen also at the level of the strange quark,
could be significant for neutrino oscillations if the radiatively
generated neutrino masses follow a pattern similar to the
massive leptons.

The up and down quarks and the electron  have no Higgs
components at all.
The scalar Higgs components (a scalar component
has no massless generators) are associated with unstable particles
and the stripping of these components from the up and down
quark will also  be required if proton stability is to be ensured.
Where the expected 5.5! for the up quark
goes we shall shortly see.

No components appear in the 4.4!=U(1) column. The two 4.4!
components and the up-quark 5.5! component get rearranged
as the following which represent a mixture of S.U.(2) and
U(1) components;

\vspace{0.5cm}

\begin{center}
\begin{tabular}{|c|c|c|c|}
\multicolumn{4}{c}{CURRENT QUARK U(1) COMPONENTS}\\\hline
quark type&`strong-charge'&generation multiplier&
strong component\\\hline
top&+&3&2.(4!-2)\\\hline
bottom&-&3&4.(4!-2)\\\hline
charm&+&2&2.(4!-2)\\\hline
strange&-&2&4.(4!-2)\\\hline
up&+&1&2.(4!-2)\\\hline
down&-&1&4.(4!-2)\\\hline
\end{tabular}
\end{center}

\vspace{0.5cm}

\noindent
The total mass represented by the
above table is exactly 5.5!+2.4.4! as required
so that every part of the Higgs mass is accounted for
except the components $3.3!+...$ the masslessness of
which induced the breakdown of the initial Higgs unit
of mass 10!$\approx$83GeV in the first place.
 These are the exotic U(1)
components we have sought to complete our calculation
of proton and neutron masses. They represent a truely
remarkable piece of structure of the Higgs field
because with them the scalar components have been 
transmuted into tetrahedral spinor units of mass
with clear mixing  implied between quark generations. 
`Higgs' units, such as the $2.9.S_9$ 
associated with the top 
appear with no gauge field corrections to
them in mass calculations (a common feature of
all Higgs components) which is an indicator that
they are `buried outside time'; they are situated
effectively `inside' the timeless space demarcated
by the boundary of tetrahedral units (one can think of 
them as occupying the inner surface of the boundary of 
the geometry whereas the gauge fields occupy the outer
surface -  the time dimension does not
exist within the boundary of any affine-set
geometry embedded in space-time). This is discussed in more
detail in \cite{filewood}. In that paper we also saw
that the current-quark U(1) components could be
interpreted in terms of a strong-interaction potential
between current quarks \cite{filewood}.
 In the case of the up  quark 
 four out of the six surfaces of the cube (each quark is
a `cube' in this theory)
are electro-magnetically charged (i.e. 2/3rds). 
The remaining  two
surfaces of the cube are uncharged electro-magnetically
but acquire a `Higgs-charge', or perhaps
more appropriately a `strong-charge' because 
it can be interpreted as a strong interaction
potential; in the case of the up
  the
two units represented in the above chart (vis-a-vis 4 for
the down) are the complement of the
four (res.2) electro-magnetically
charged square surfaces of the cube;
accounting for the total of six square
surface sub-geometries for each cube.
One can easily see this structure in the
quark operators given in the previous section.
 Presumably this means that, for example
in the case of the up, only four of the
six surfaces of the cube are being presented to 
space-time; the other two surfaces are outside time
(?folded in).

The `Higgs-charge' concept arose because the above U(1)
mass units in question have been given massless generators
and in space-time this would mean electro-magnetic
charge due the the spinning triangular sub-geometries.
The Higgs charge seems to be an analogue of electro-magnetism
but in a timeless colour space (it might lend support to
the idea of instantons?). From a phenomenological
point of view the Higgs charges are difficult to understand
(will they effect the results of deep inelastic scattering
at ultra-high energy in spite of the fact that they are
located in a timeless 3-space? Can they be `pushed' into
space-time at high energy leading to deviation from the
standard model predictions at high energy?)
but the methodology required to use them in calculations
is known precisely. To calculate the energy associated with
 current quark U(1) components  one averages
the absolute value of the sum of the respective `Higgs charges'.
For example, in the case of two up quarks the value is;

\begin{equation}
\mbox{up-up current quark Higgs energy}={1\over2}|+2.(4!-2)+2.(4!-2)|
=2.(4!-2)
\end{equation}
and so-on for the other components. One sees that the
choice of charge is arbitrary except that the up-type
quarks and down-type quarks must have opposite sign
just as they do for conventional electro-magnetic charge.
The logical choice is to give the same Higgs charge 
sign as the E.M. charge for each quark. Thus for the
up-quark four of the six surfaces have positive 
electro-magnetic charge (2/3rds of the surface of the
cube) and two of the six surfaces (1/3rd of the surface)
have a positive Higgs charge.
It is these terms which
define the additional masses for the proton and neutron
expressions set earlier in the paper.  The
proton we have three kinds of couplings; one up-up
and two up-downs so the sum $\Sigma = 4(4!-2)$ and for the neutron we have
two up-downs and one down-down $\Sigma=6.(4!-2)$. 

Notice that, from the point of view of the
Higgs field,  it really is only at the level of the U(1)
Higgs charges that the nucleons can be considered
to have any component of the strange or other
higher-generation quark content since these Higgs
energies may alternatively be interpreted as
being generated fractionally by the first generation
component of the higher generation quarks - which
are identical to that of the up and down
(the pattern is repeated identically for each
generation so that, for example, a top-quark
has a +2(4!-2) corresponding to a first, second
and third generation representation -
although what you get depends upon the actual
quark content of a given hadron).
 We will later see however
that the remaining components of the current up and down
quarks (developed in the preceding section) 
have a similar  interpretation in terms of mixing 
with higher generations not obvious from the 
previous calculations performed for the nucleons
\cite{filewood}.

\section{Ground state meson masses}

 Now we start on the main purpose of this paper;
a study of ground state meson masses
in order to gain insight into the 
mass structure of higher generation
quarks.

The nucleon masses and the electron mass
are particularly simple
which  might be expected for the fundamental stable
matter constituents of the universe. They were examined
closely in \cite{filewood}. However
we can use the same techniques of analysis to
determine ground state meson masses. The techniques are
a generalisation of the methods already developed for
the nucleons with slight modifications
because of the presence of only two effective
quarks in a meson. 

For the nucleons we had only first generation
quarks to consider and we developed a three-colour
particle vector on which the current quark representations
acted. For mesons the situation is slightly
different because in general only one
colour is active in the particle vector.
For example, we may consider that 
a $\pi^{+}$ meson consists of the quark content
$U_r\bar{D}_r$  where $r$ is a single colour.
In calculating the corresponding constituent mass
this means that only one of the possible three
colour components of the particle vector becomes
massive and the remaining two are
treated as  massless. However, we then {\it{sum}}
over all possible permutations of discrete colour
which gives a multiplier of 3. (By contrast with 
the nucleons we had three massive components and
hence a multiplier of 3! for the colour permutations).

For a first generation particle vector representation
we thus have only one massive component but in two 
possible parity states over which we must sum;

\vspace{0.5cm}

\begin{equation}
\left(
\begin{array}{cc}
c_i&0\\
0&\bar{c}_i
\end{array}
\right)
\;\;\;\;\;\;\;\;\;\;\;\;
\left(
\begin{array}{cc}
\bar{c}_i&0\\
0&{c}_i
\end{array}
\right)
\label{paritystates}
\end{equation}

\vspace{0.5cm}

The matrix order of $\bar{C}_i$ is $(4!-2)$
and the matrix order of $C_i$ is simply $4!$
(the generators of the $T_c$ group are in the
barred matrix and assumed to be massless intrinsic
spin generators in the transition to a field theory).
Current quark
coupling to both possible parity combinations 
introduces a multiplier of 2. Thus the
total constituent order for a meson with two
first generation quarks becomes;

\vspace{0.8cm}

\centerline
{\epsfig{file=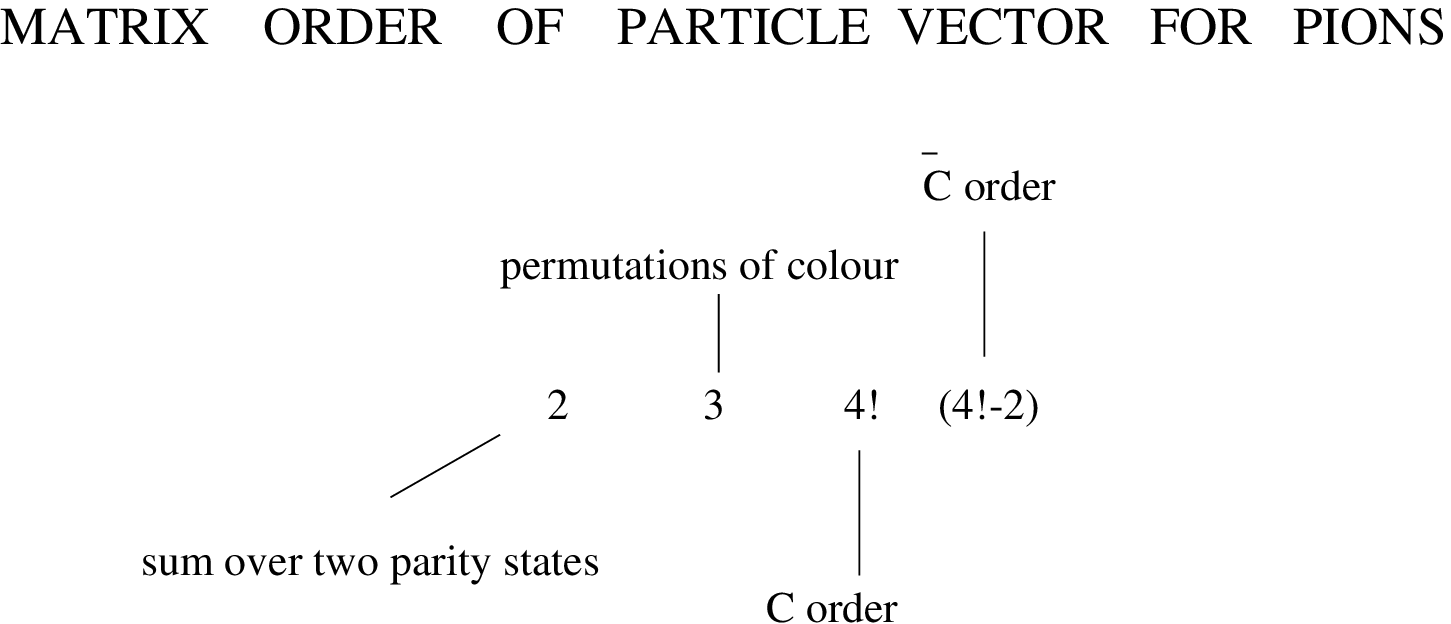,width=10cm}}

\vspace{0.8cm}

The order of the constituent gluons is similar.
Typical gluon formats are;

\vspace{0.5cm}

\begin{eqnarray}
C_i&\left(
\begin{array}{cc}
q_iq_j^*&0\\
0&I_3
\end{array}
\right)
C_j^*
\;\;\;\;\;;\;\;\;\;\;
C_i\left(
\begin{array}{cc}
q_iq_j^*&0\\
0&{\cal{I}}_k
\end{array}
\right)
C_j^*
\nonumber
\\
\;&\nonumber
\\
C_i&\left(
\begin{array}{cc}
q_i{\cal{I}}_kq_j^*&0\\
0&I_3
\end{array}
\right)
C_j^*
\;\;\;\;\;;\;\;\;\;\;
C_i\left(
\begin{array}{cc}
q_i{\cal{I}}_kq_j^*&0\\
0&{\cal{I}}_k
\end{array}
\right)
C_j^*
\end{eqnarray}

\vspace{0.5cm}

\noindent
and the order of any gluon operator is
the same as the order of the product $C_iC_j^*$
which is $4!^2$ (note that $^*$ indicates the
 conjugate transpose  matrix which is not the same
as the barred matrix). With discrete colour there
are two possible transitions for any given
colour state i.e.;

\[
C_i
\rightarrow
C_j\;\;\;\;
\mbox{and}\;\;\;\;
C_i
\rightarrow
C_k
\]

\noindent 
where \{i,j,k\} are the three possible discrete
colours. This  doubles the gluon matrix order.
There is a further doubling of the order 
due to coupling to either of the two possible
parity states of the particle vector which gives
the glue matrix order for a ground state meson
with only one pair of first generation quarks
as $2.2.(4!)^2$.

 When dealing with higher generation quarks the
situation is similar  but more complex;
 generally the rule holds that second generation
quarks have double representations and third generation
quarks have triple representations. Notice that the 
U(1) Higgs components have exactly this structure.

The rules for meson calculation are as follows;

1; From the possible matrix representations of
a particle vector choose a representation capable
of expressing the quark generation content of 
the meson. This is the G or generation number $1{\leq}G{\leq}3$
where G is the highest generation present. Thus for example
a bottom containing meson  will have a G=3. A $K^0$ meson will
have G=2 etc.

2; Calculate the P or parity number which represents the
number of parity formats present in the particle vector.
(For example, in the case of the first generation we
only have two formats; left and right given by 
eq.(\ref{paritystates});
in higher generation representations there can be more than
two!). Parity `formats' are not always the
same as conventional  parity; 
only for first-generation mesons
is there a true correspondence. The meaning of this will
become clear as we proceed through examples.

3;
Calculate the number S. S is the number of 
distinct states seen by the gluons and is a multiplier
for the gluon energy. S stands for `summed quantum
numbers' and is sometimes represented as $\Sigma$
in the text.

4;
Calculate the representation number R which is the
total number of representations needed to represent
the quark content  of the meson.
 Up-type quarks have two representations
in the second and third generation components. These
additional representations eliminate spinorial identities
which otherwise appear in the charm or top
in the higher generation content. We will discuss this
in more detail as we come to it. These additional representations
do {\it{not}} contribute to the generation number G. However,
these additional representations induce additional permutations
of `parity format' and do change the P number. 
Such permutations are
called {\it{superparity}} in the text because of the induced
interchange of scalar and spinorial components.

5;
Calculate the `bare' constituent quark mass $=G.P.22.24.3$
for three possible colours.

6;
Calculate the constituent `bare' gluon mass
$=R.S.24.24.2$ for interchange of 
one pair of colours.

7; Sum the terms for the charged current
quark mass = C using the standard format
of current quark representations.

8;
Add the universal radiative correction multiplier
${\cal{R}}$
for quark-lepton unification;

=${\cal{R}}.(G.P.22.24.3+R.S.24^2.2+C)$

9;
Add on the appropriate scalar and U(1)
Higgs field components for the quark content
of the meson; these have no radiative correction.

This technique should work for any meson. We have
already seen that it works for the nucleons;
modulo changes in the formula for the presence
of three discrete colours `simultaneously'
in the particle vector with  the corresponding
glue order changes for three colours of quark
simultaneously present. We will
now examine some specific examples to illustrate
the above principles. We will see that we get quite
good results except for the charmed-bottom meson
where the is a marked deviation away from the
empirical result.

\section{The pions}

The charged pion consists of an up/antidown or
down/anti-up combination. The calculation is the
same for both. We calculate first the {\it{constituent}}
mass which consists of both quark and gluon energy. These
values are found in the theory by summing over the matrix
order of the representation of the particle vector and 
glue. Note that the particle vector does not specify
the quantum state of the object but sums over all
possible discrete quantum states. The specific quantum
numbers of the state are specified by the {\it{current quark}} 
components which are calculated separately.

For the pions G=1, P=2 so the `bare' constituent quark
order is 2.3.22.24=3168.

S is best calculated using a table.

\begin{center}
\begin{tabular}{|c|c|c|}
\multicolumn{3}{c}{Pion Structure}
\\\hline
Parity combinations&$\pm$&$\mp$\\\hline
First generation parity&-1&+1\\\hline
Superparity&0&0\\\hline
Summed numbers&-1&+1\\\hline
\end{tabular}
\end{center}

A word of explanation here. The two parity combinations
are the same as in eq.(\ref{paritystates}). For example we may
regard the $\pm$ as the diag.\{$C,\bar{C}$\}
 combination and the
$\mp$ as diag.\{$\bar{C},C$\} matrix. It makes no difference
to the mass calculation which we regard as the left-handed
part of the quark particle vector and which the right-handed
but we routinely adhere to the pattern $\pm=-1$
and $\mp=+1$ for first generation parity.
The superparity
of the two states represented here is identical
and both have been set at zero (superparity is only
formally definable for mesons with 2nd and 3rd
generation valance quarks; we will discuss this when we reach
it).
 The summed numbers are simply the sum of the superparity
and parity numbers
in the case at hand 
(but whith higher generation quarks includes
other quantum numbers). 
S is the number of {\it{different}} summed
numbers $\Sigma$ and represents the number of distinct states
coupled to  by gluons. For example, in the case of the 
pions the possible summed numbers are +1 and -1  so that S=2 
(physically this is equivalent
to the statement that the strong interaction is independent
of the parity of the state and couples to both right and
left-handed states doubling the gluon matrix order).

We only need one representation for the pions so 
R=1. (R is the number of $\pm$'s and or $\mp$'s  per box in the
parity combinations section of the table -
for the pions each box has only one such symbol).

Thus the gluon order is $R.S.2.24^2=2304$.

To calculate the current quark mass we need
a standard representation of the quark content. The notation
is identical to that used for the nucleons and
we create an anti-down (or anti-up) by taking the
complex-conjugate representation. For a meson we 
have colour and anticolour combinations of quarks
in the current-quark content. The red-antired $\pi^+$
content is;

\begin{center}
\begin{tabular}{ccc}
\multicolumn{3}{c}{Charged Pion Current Quarks.}
\\\hline\\
&$\;\;\;\;
$strong component$\;\;\;\;$&E.M. component\\
\\
red$\rightarrow$&${\cal{I}}\;\;\;\;\;\;\;\;\;\;q$
&$I\;\;\;\;\;\;\;\;\;\;q*$\\
green$\rightarrow$&$q*\;\;\;\;\;\;\;I$&$q*\;\;\;\;\;\;\;I$\\
blue$\rightarrow$&$q*\;\;\;\;\;\;\;I$&$q*\;\;\;\;\;\;\;I$\\\\
&$\;\;\;\;\;{\stackrel{\uparrow}{\mbox{up}}}\;
\;{\stackrel{\uparrow}
{\mbox{anti-down}}}$
&$\;\;\;\;\;{\stackrel{\uparrow}{\mbox{up}}}\;
\;{\stackrel{\uparrow}
{\mbox{anti-down}}}$\\\\\hline
\end{tabular}
\end{center}

Notice that the definition of unit charge differs with
a bosonic meson to a fermion. The application of the 
above current quarks to the standard particle vector
interchanges the $C$ and $\bar{C}$ matrices but adds
a + sign to one and a - sign to the other which is
the definition of unit charge for the boson. (Incidentally
this notation means that only one neutral pi meson
occurs in the representation along with the two charged
states; for this reason we will call this
method of defining unit charge the `adjoint' 
representation as distinct to the `fundamental'
representation used for the fermionic nucleons).
Note, however, 
 that the bosonic nature of the object is represented
in its {\it{current-quark}} content not in the particle
vector.

The calculation of current-quark mass proceeds as for the
case of the nucleons. We need only sum one colour combination
of current quarks since the particle vector includes all
possible colour permutations. The masses are read off the
chart. The three q*'s in the E.M. component generate,
with the doubling due to the coupling to the second
parity state (as per the nucleons), 6.(4!-2) units and,
 with the two cancellations
in the strong component
(the q* cancels the q and the ${\cal{I}}$
cancels the I in the strong matrix),
 the residual q* and 4 identities
(one from the strong component and three from the
EM part) generate,
again with parity doubling,
5.2.4!. Summing all these components and applying rule 8
we obtain, with some reorganisation,
 ${\cal{R}}.\{7!+6!+4.4!-12\}$ for the total
mass.
To this we must add the Higgs U(1) components. This is
quite simple in the case of the charged pion and was
previously calculated as (4!-2) for an up-down type
combination with no radiative correction.
However the resulting mass is about 136MeV whereas the
empirical mass is 139.5679MeV. Comparison with 
empirical data shows that the deficiency is
almost exactly a scalar Higgs of value 5!+4! which suggests
that there is a component of strange quark present
(which has an identical first-generation current-quark
representation to the down except for the Higgs 
scalar component). The effective  component of strange quark
present is given by the ratio;

\begin{equation}
{\mbox{{Higgs scalar in pion}}\over{\mbox{Higgs scalar in strange}}}=
{{5!+4!}\over{6!+3.5!}}
\end{equation}

\noindent
where the strange Higgs scalar component has been taken
from the Higgs component table. Thus if we use a mixture 
of down and strange to form the charged pion
we obtain the mass $\pi^{\pm}=139.5679$MeV. which is 
close to the
correct value of 139.57018(35)MeV{\cite{groom}}.

The calculation for the neutral pion proceeds in an
analogous fashion with the appropriate current-quark
content substituted. The Higgs U(1) component is
most accurately given by averaging one up-up,
 one up-down and one down-down component from the
Higgs U(1) table and the result of the
calculation is $\pi_0\approx
134.9728$MeV. which is the correct value{\cite{groom}}.
This value, however, does {\it{not}}
include any component of Higgs scalar for
the strange quark which appears to be absent
in the neutral pion. Why a component of strange
appears explicitly in the $\pi^{\pm}$ and not
in the $\pi^0$ is unclear. It is possible that
the strange-down mix (Cabibbo-type)
is an incorrect interpretation
and that the presence of singlet SU(2) and U(1)
Higgs components in the charged pions represents
an incomplete formation of the Higgs U(1) 
components with the $\pi^{\pm}$'s; that is to say,
that the up-quark has a singlet SU(2) and U(1) scalar
piece in the charged pion.
An answer will require a more sophisticated 
version of the theory 
of the Higgs field. Interestingly the charged B meson
seems to have identical Higgs SU(2) and U(1) singlets
associated with the up quark component (see subsequent 
calculation) which are also missing in the neutral B
meson.

\section{Higher generation quarks.}

For the up and down species of quarks
we employed a block diagonal representation
of both the particle vector and  the current
quark operators that act on the particle
vector. Thus for the $T_{c_i}$ group each colour
$6\times6$ block has the format;

\[
\left(
\begin{array}{cc}
C_i&0\\
0&\bar{C}_i
\end{array}
\right)
\;\;\;or\;\;\;
\left(
\begin{array}{cc}
\bar{C}_i&0\\
0&{C}_i
\end{array}
\right)
\]

\noindent
where $C_i$ is a  $T_c$ colour matrix of 
colour i.
This is not the only way that a
consistent representation can be formed capable of 
coupling to a given current quark operator
representation however. We will call this arrangement,
when the $q_i$'s of the current quark 
operators and the $C_i$'s 
of the particle vector are in the same format
as `in phase'. It only applies to the situation
where a hadron has first generation quarks.
Consider the following matrix;

\vspace{0.5cm}

\[
\left(
\begin{array}{cccccc}
\lambda_{11}&0&0&0&\lambda_{12}&0\\
0&\bar{\lambda}_{11}&0&\bar{\lambda}_{12}
&0&0\\
0&0&-1&0&0&0\\
0&\bar{\lambda}_{21}&0&\bar{\lambda}_{22}&0&0\\
\lambda_{21}&0&0&0&\lambda_{22}&0\\
0&0&0&0&0&+1
\end{array}
\right)
\]

\vspace{0.5cm}
\noindent
where the $\lambda_{ij}$ are the individual
elements of a given $C$ matrix.
The above matrix is a higher generation representation;
it is blue format following the standard used in this
paper and the colour is dictated by the position of 
the identities on the main diagonal. We call this a
`phase-shifted' representation. Now if we apply a 
similarly `phase-shifted' blue quark operator the
physical results are unaltered. However if we operate
on the phase-shifted particle rep. with the original
$q_b$ $6\times6$ matrix (in this example a blue
down quark q operator);

\vspace{0.5cm}

\[
\left(
\begin{array}{cccccc}
i&0&0&0&0&0\\
0&i&0&0
&0&0\\
0&0&-1&0&0&0\\
0&0&0&-i&0&0\\
0&0&0&0&-i&0\\
0&0&0&0&0&-1
\end{array}
\right)
\]

\vspace{0.5cm}

\noindent
we get the following;

\vspace{0.5cm}

\[
\left(
\begin{array}{cccccc}
\lambda_{11}&0&0&0&\lambda_{12}&0\\
0&\bar{\lambda}_{11}&0&\bar{\lambda}_{12}
&0&0\\
0&0&+1&0&0&0\\
0&\bar{\lambda}_{21}&0&\bar{\lambda}_{22}&0&0\\
\lambda_{21}&0&0&0&\lambda_{22}&0\\
0&0&0&0&0&-1
\end{array}
\right)
\]

\vspace{0.5cm}

\noindent
which is the same as the original matrix 
in terms of the positions of the $C$
and $\bar{C}$ matrices (note that the 
actual entries $C_{ij}$ or  $\bar{C}_{kl}$
 will have changed
in value - we are only interested in whether
the barred and unbarred matrices mix or not)
but the sign on the identities on the main diagonal
has changed. To maintain consistency we modify the
down blue q operator by changing sign on the
operator identities (compare with previous);

\vspace{0.5cm}

\[
\left(
\begin{array}{cccccc}
i&0&0&0&0&0\\
0&i&0&0
&0&0\\
0&0&+1&0&0&0\\
0&0&0&-i&0&0\\
0&0&0&0&-i&0\\
0&0&0&0&0&+1
\end{array}
\right)
\]

\vspace{0.5cm}

\noindent
This sign change has three effects;

1. it distinguishes the first and 
higher generation quark operators,

2. it  causes the higher generation q operator
to act like a script identity operator which 
un-couples it from the $T_c$ generators; it
behaves like a scalar rather than a spinor 
operator. As we shall see, this actually has
an effect on particle mass calculation
which may represent the first evidence of 
supersymmetry in nature and

3. it is required to maintain charge conservation.

Remarkably, as can be checked, the 
(non phase-shifted) script identity
operator requires a similar change
in the sign of its' component which couples
to the diagonal identity of the $C$ or $\bar{C}$
matrix which causes it to act as a discrete spinor
generator! Thus under the altered phase
representation of the particle vector the
non-phase shifted current quark operators undergo
an interchange of scalar and fermionic pieces!

These changes apply to all second and third generation
representations. The altered phase couplings can
be pictorially represented;

\vspace{0.6cm}

\centerline
{\epsfig{file=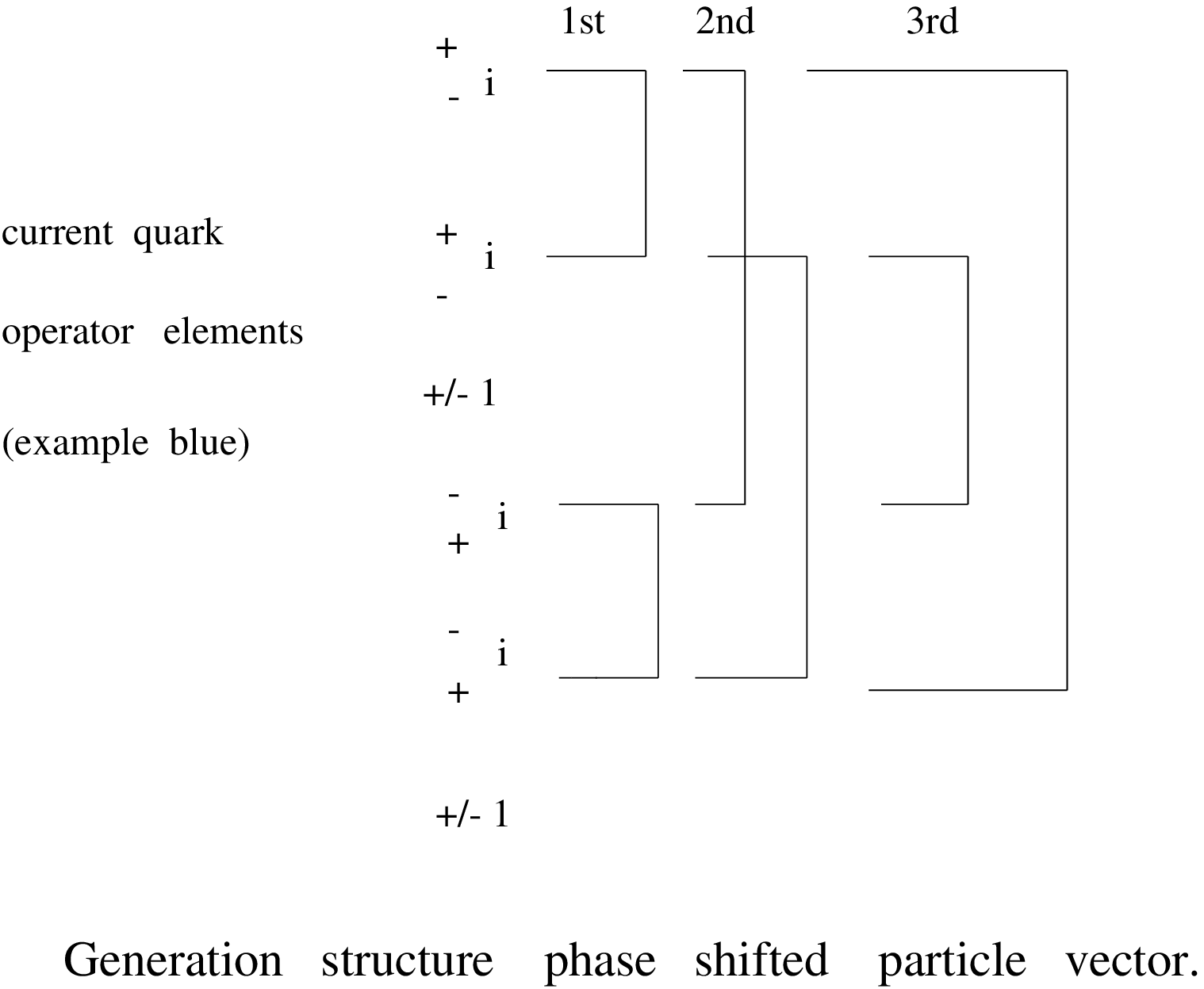,width=10cm}}

\vspace{0.6cm}

\noindent
In the above diagram the diagonal entries in
a q operator (here blue) have been listed
in the  verticle column of entries. The first generation
couplings are the in-phase ones and the second and
third generation couplings are out of phase.
The `coupling' lines in this diagram represent the
relative position in the 6x6 matrix occupied
by the $C_i$ and $\bar{C}_i$ with respect to the
discrete current-quark operator on the left.
There is an apparent  ambiguity  between what we call
second and what we call the third generation
representation; it makes no difference 
in practice since
the third generation is found by summing over
all three possibilities. The first generation is
always fixed in phase to that of the current
quark operators; in the notation used here the
`phase' of the operators stays fixed and that
of the particle vector changes between different
generation representations. It might be possible 
to represent the symmetry using fixed particle 
vector and changing the phase of the operators
but the physics will be the same.

Note two interesting things;

1; the operator structure undergoes a 
supersymmetry transformation in the
higher generation representations and

2; the representation theory limits the
number of possible generations to three;
this seems to be a unique feature of the
discretised form of QCD since the continuum
theory does not demand only three generations.

As a general interpretation of the higher 
generation representations probably the simplest
way to consider them is as additional possible ways
in which the symmetry can be expressed geometrically.
The `in-phase' representation dominates the first
generation representation because this is the representation
in which electro-magnetic charge is defined in a 
consistent way.

\section{Kaons}

In this section we calculate
the mass of the spin-0 $K^0$ and $K^{\pm}$
using the principles outlined in the previous
sections.

The basic principles are as with
the pions. A particle vector is established
capable of expression all permutations of
particle type and the actual quark content
is carried in the current quarks. Thus
the `stangeness' quantum number, which will
be $\pm1$, is carried by the current quark
content but the particle vector carries
the potential for either strangeness quantum
number. The G number is clearly 2 since
the highest generation number in a Kaon
corresponds to the strange quark. The particle vector
contains a representation for the first generation
and a second one for the second generation; it is
duplicated. The tabulation for the structure
of the particle vector is as follows;

\vspace{0.5cm}

\begin{center}
\begin{tabular}{|c|c|c|c|c|c|c|c|c|}
\multicolumn{9}{c}{Kaon Structure;
Particle Vector}
\\\hline
$C/{\bar{C}}$ combinations (1st,2nd)
&\multicolumn{2}{c|}{$\pm\;\;\pm$}
&\multicolumn{2}{c|}{$\pm\;\;\mp$}
&\multicolumn{2}{c|}{$\mp\;\;\pm$}
&\multicolumn{2}{c|}{$\mp\;\;\mp$}
\\\hline
(First generation) parity&-1&$\;$&-1
&$\;$&+1&$\;$&+1&$\;$\\\hline
(Second generation) strangeness&$\;$&+1&$\;$&-1
&$\;$&+1&$\;$&-1\\\hline
Superparity&\multicolumn{2}{c|}{+2}
&\multicolumn{2}{c|}{-2}
&\multicolumn{2}{c|}{-2}
&\multicolumn{2}{c|}{+2}
\\\hline
Summed Numbers&\multicolumn{2}{c|}{+2}
&\multicolumn{2}{c|}{-4}
&\multicolumn{2}{c|}{0}
&\multicolumn{2}{c|}{+2}
\\\hline
\end{tabular}
\end{center}

\vspace{0.5cm}

From which we can read 
from the first row that R=2
(the number of sets of $\pm$ signs per
box in the first row) and  P=4. 
P is the number of different boxes in the
first row;
in this case the `parity' of the second
generation component, which is the strangeness
quantum number, contributes to the total
number of different `parity' states -
the `stangeness quantum number' is literally
the parity of the second component of the
particle vector; more specifically it is 
related to the corresponding part of the 
current quark that couples to the second
generation component of the particle vector. 
Strangeness is actually defined in the opposite
way to the `parity' in the first generation;
for example here $\pm$ has been chosen as
odd parity in the first generation component whilst
$\pm$ has been chosen as even for the strangeness.
The `superparity' depends on whether the combination
of parity and strangeness assignments is odd or
even. Thus, for example $\pm\pm$ is even with 
value +2, whilst $\pm\mp$ is odd with value -2. The
S number is the number of different summed numbers
and represents the number of distinct states 
recognised by gluons. A summed number is formed by
adding all the entries above it in a given column.
In the present example
S=3.

We can immediately write down the particle vector
order as $G.P.3.22.24=12672$. The gluon order
is $R.S.2.24.24=6912$.

The components we have just calculated represent
the constituent quark content of the Kaons;
they are common to the charged and neutral versions
and represent a sum over all possible discrete states
that the quark pair can exist in.
We must now determine the current quark mass
of the different Kaons. From the Higgs mass
chart we know that the strange quark has
Higgs scalar contributions 6!+3.5! with no
gauge field radiative corrections. To calculate the
remaining components of mass we need a chart of
current quark components. In the second generation
representation only the strange quark is represented;

\begin{center}
\begin{tabular}{ccc}
\multicolumn{3}{c}{Neutral kaon; Current Quarks
(First generation).}
\\\hline\\
&$\;\;\;\;
$strong component$\;\;\;\;$&E.M. component\\
\\
red$\rightarrow$&$q*\;\;\;\;\;\;\;\;q$
&$q\;\;\;\;\;\;\;\;\;\;\;q*$\\
green$\rightarrow$&${{I}}\;\;\;\;\;\;\;\;\;\;I$
&$I\;\;\;\;\;\;\;\;\;\;\;\;I$\\
blue$\rightarrow$&${{I}}\;\;\;\;\;\;\;\;\;\;I$
&$I\;\;\;\;\;\;\;\;\;\;\;\;I$\\\\
&$\;\;\;\;\;{\stackrel{\uparrow}{\mbox{down}}}\;
\;{\stackrel{\uparrow}
{\mbox{anti-strange}}}$
&$\;\;\;\;\;{\stackrel{\uparrow}{\mbox{down}}}\;
\;{\stackrel{\uparrow}
{\mbox{anti-strange}}}$\\\\\hline
\end{tabular}
\end{center}
\begin{center}
\begin{tabular}{ccc}
\multicolumn{3}{c}{Neutral kaon; Current Quarks
(Second generation).}
\\\hline\\
&$\;\;\;\;
$strong component$\;\;\;\;$&E.M. component\\
\\
red$\rightarrow$&$\;\;\;\;\;\;\;\;q$
&$\;\;\;\;\;\;\;\;\;\;\;q*$\\
green$\rightarrow$&$\;\;\;\;\;\;\;\;\;\;I$
&$\;\;\;\;\;\;\;\;\;\;\;\;I$\\
blue$\rightarrow$&$\;\;\;\;\;\;\;\;\;\;I$
&$\;\;\;\;\;\;\;\;\;\;\;\;I$\\\\
&$\;\;\;\;\;\;
\;{\stackrel{\uparrow}
{\mbox{anti-strange}}}$
&$\;\;\;\;\;\;\;\;\;
\;{\stackrel{\uparrow}
{\mbox{anti-strange}}}$\\\\\hline
\end{tabular}
\end{center}

\noindent
The q's cancel the q*'s in the first generation 
representation of the current quarks leaving 
the eight identities which, with the parity
doubling (coupling to both the right and left
handed components of the particle vector)
we obtain a mass of 16.24.

For the second generation representation
there is the analogue of a supersymmetry
transformation on the components with the
identities in the E.M. component coupling to
generators (i.e. (4!-2) unit which is explicitly spinorial)
and the q's and q*'s coupling to the
scalar value 4!; this is the reverse of the 
pattern seen in the first generation components.
This kind of transformation is essential to
eliminate any electro-magnetic charge generation
component of the second generation representation;
since $I^{\dagger}=I$ the identities coupled to the
generators generate no charge. (It is interesting to note
that the existence of this kind of structure may be
the first evidence of supersymmetry in nature -
it is a form of concealed supersymmetry and, of
course, in the present example discretised and not
continuous).
The q therefore cancels the q* in the second 
generation representation and the mass of
the second generation representation,
again with doubling for coupling to both
parity components, is 4.4!+4.(4!-2).

Finally we must calculate the Higgs U(1)
components from the table. The strange has
a component of 4(4!-2) for each of its 
two generation components. There is an
`interaction' between the down and each of the
two components of the strange leading to a doubling
of the down-down U(1) order to 8(4!-2).

Summing all these components and applying
the universal radiative correction ${\cal{R}}$
we obtain;
\begin{eqnarray}
K^0{\mbox{mass}}&=&
{\cal{R}}.\{
\mbox{Particle order + gluon order + current quarks\}
+ Higgs sector}
\nonumber\\
&=&
{\cal{R}}.\{
12672+6912+568\}+6!+3.5!+8.(4!-2)
\end{eqnarray}

\noindent
which translates into a mass of $K^0= 497.0368$MeV
using the electron rest mass expression which compares with
the empirical value of 497.672(31)MeV.

For the charged Kaon we must alter the current
quark content to accommodate the up or anti-up
quark. The appropriate chart is;

\begin{center}
\begin{tabular}{ccc}
\multicolumn{3}{c}{Charged kaon; Current Quarks
(First generation).}
\\\hline\\
&$\;\;\;\;
$strong component$\;\;\;\;$&E.M. component\\
\\
red$\rightarrow$&${\cal{I}}\;\;\;\;\;\;\;\;\;\;\;q$
&$I\;\;\;\;\;\;\;\;\;\;\;\;\;q*$\\
green$\rightarrow$&$q*\;\;\;\;\;\;\;\;\;\;I$
&$q*\;\;\;\;\;\;\;\;\;\;\;\;I$\\
blue$\rightarrow$&$q*\;\;\;\;\;\;\;\;\;\;I$
&$q*\;\;\;\;\;\;\;\;\;\;\;\;I$\\\\
&$\;\;\;\;\;{\stackrel{\uparrow}{\mbox{up}}}\;
\;{\stackrel{\uparrow}
{\mbox{anti-strange}}}$
&$\;\;\;\;\;{\stackrel{\uparrow}{\mbox{up}}}\;
\;{\stackrel{\uparrow}
{\mbox{anti-strange}}}$\\\\\hline
\end{tabular}
\end{center}
\begin{center}
\begin{tabular}{ccc}
\multicolumn{3}{c}{Neutral kaon; Current Quarks
(Second generation).}
\\\hline\\
&$\;\;\;\;
$strong component$\;\;\;\;$&E.M. component\\
\\
red$\rightarrow$&$\;\;\;\;\;\;\;\;q$
&$\;\;\;\;\;\;\;\;\;\;\;q*$\\
green$\rightarrow$&$\;\;\;\;\;\;\;\;\;\;I$
&$\;\;\;\;\;\;\;\;\;\;\;\;I$\\
blue$\rightarrow$&$\;\;\;\;\;\;\;\;\;\;I$
&$\;\;\;\;\;\;\;\;\;\;\;\;I$\\\\
&$\;\;\;\;\;\;
\;{\stackrel{\uparrow}
{\mbox{anti-strange}}}$
&$\;\;\;\;\;\;\;\;\;
\;{\stackrel{\uparrow}
{\mbox{anti-strange}}}$\\\\\hline
\end{tabular}
\end{center}

The calculation proceeds as before. The particle
vector and gluon calculations are identical. The mass
of the first generation chart can be read off
as 2.\{3.(4!-2)+5.4!\}. The second generation
component is identical to the neutral Kaon and
is 2.\{2.4!+2.(4!-2)\}. The Higgs U(1) component
is twice the up-down value and is therefore 2.(4!-2).
Thus the mass is;
\begin{equation}
K^{\pm}{\mbox{mass}}={\cal{R}}\{
12672+6912+556\}+6!+3.5!+2.(4!-2)=493.7143MeV
\end{equation}
\noindent
which compares favourably with the empirical value of
493.677(13)MeV{\cite{groom}}.

\section{D mesons}

In this next example we will study the simplest
mesons containing charm to see how the 
second generation structure of the strange is 
extended.

Charm quarks, as for the case of the
strange quark, have a second generation
representation but it is dual. The reason
for this seems to be the presence of the
script identity. Since ${\cal{I}}^{\dagger}
\neq{\cal{I}}$ a supersymmetry transformation
from the first to second generation component
produces an effective additional charge.
This is eliminated by duplicating the second
generation representation in the particle vector.
A working chart for the particle vector and gluon
contributions to $D^{\pm}$ and $D^0$ is;

\begin{center}
\begin{tabular}{|c|c|c|c|c|c|c|c|c|c|c|c|c|}
\multicolumn{13}{c}{D meson Structure;
Particle Vector}
\\\hline
&\multicolumn{2}{c|}{$\mp\;\;\;\;\mp\mp$}
&\multicolumn{2}{c|}{$\mp\;\;\;\;\mp\pm$}
&\multicolumn{2}{c|}{$\mp\;\;\;\;\pm\pm$}
&\multicolumn{2}{c|}{$\pm\;\;\;\;\mp\mp$}
&\multicolumn{2}{c|}{$\pm\;\;\;\;\pm\mp$}
&\multicolumn{2}{c|}{$\pm\;\;\;\;\pm\pm$}
\\$C/{\bar{C}}$  (1st,2nd)
&\multicolumn{2}{c|}{$\;$}&
\multicolumn{2}{c|}{$\mp\;\;\;\;\pm\mp$}
&\multicolumn{2}{c|}{$\;$}&\multicolumn{2}{c|}{$\;$}
&\multicolumn{2}{c|}{$\pm\;\;\;\;\mp\pm$}
&\multicolumn{2}{c|}{$\;$}
\\\hline
 parity&+1&$\;$&+1
&$\;$&+1&$\;$&-1&$\;$&-1
&$\;$&-1&$\;\;$\\\hline
 charm&$\;$&+1&$\;$&-1
&$\;$&+1&$\;$&+1&$\;$&-1&$\;$&+1\\\hline
Sup.parity&\multicolumn{2}{c|}{+2}
&\multicolumn{2}{c|}{-1}
&\multicolumn{2}{c|}{-2}
&\multicolumn{2}{c|}{-2}
&\multicolumn{2}{c|}{+1}
&\multicolumn{2}{c|}{+2}
\\\hline
$\Sigma$ &\multicolumn{2}{c|}{+4}
&\multicolumn{2}{c|}{-1}
&\multicolumn{2}{c|}{0}
&\multicolumn{2}{c|}{-2}
&\multicolumn{2}{c|}{-1}
&\multicolumn{2}{c|}{+2}
\\\hline
\end{tabular}
\end{center}

\noindent
From which we can read
that G=2 (contains a second generation quark),
P=8 (total number of permutations of the
plus/minus signs),
R=3 (the number of plus/minus signs per 
combination) and S=5 (the number of different summed
numbers). Plugging these into the mass formula
gives a total order for the particle vector
plus glue as 42624.

The next step is to evaluate the current quark
mass. Let us consider the neutral D meson first.

\begin{center}
\begin{tabular}{ccc}
\multicolumn{3}{c}{$D^0\;\;\;$; Current Quarks
(First generation).}
\\\hline\\
&$\;\;\;\;
$strong component$\;\;\;\;$&E.M. component\\
\\
red$\rightarrow$&${\cal{I}}\;\;\;\;\;\;\;\;\;\;\;{\cal{I}}$
&$I\;\;\;\;\;\;\;\;\;\;\;\;\;I$\\
green$\rightarrow$&$q*\;\;\;\;\;\;\;\;\;\;q$
&$q*\;\;\;\;\;\;\;\;\;\;\;\;q$\\
blue$\rightarrow$&$q*\;\;\;\;\;\;\;\;\;\;q$
&$q*\;\;\;\;\;\;\;\;\;\;\;\;q$\\\\
&$\;\;\;\;\;{\stackrel{\uparrow}{\mbox{up}}}\;
\;{\stackrel{\uparrow}
{\mbox{anti-charm}}}$
&$\;\;\;\;\;{\stackrel{\uparrow}{\mbox{up}}}\;
\;{\stackrel{\uparrow}
{\mbox{anti-charm}}}$\\\\\hline
\end{tabular}
\end{center}
\begin{center}
\begin{tabular}{ccc}
\multicolumn{3}{c}{$D^0\;\;\;$; Current Quarks
(Second generation).}
\\\hline\\
&$\;\;\;\;
$strong component$\;\;\;\;$&E.M. component\\
\\
red$\rightarrow$&$\;\;\;\;\;\;\;\;{\cal{I}}$
&$\;\;\;\;\;\;\;\;\;\;\;I$\\
green$\rightarrow$&$\;\;\;\;\;\;\;\;\;\;q$
&$\;\;\;\;\;\;\;\;\;\;\;\;q$\\
blue$\rightarrow$&$\;\;\;\;\;\;\;\;\;\;q$
&$\;\;\;\;\;\;\;\;\;\;\;\;q$\\\\
&$\;\;\;\;\;\;
\;{\stackrel{\uparrow}
{\mbox{anti-charm}}}$
&$\;\;\;\;\;\;\;\;\;
\;{\stackrel{\uparrow}
{\mbox{anti-charm}}}$\\\\\hline
\end{tabular}
\end{center}

In evaluating the mass it is assumed that
the script identity generates a mass component
with the opposite sign to that of the identity
in the second generation representation. With the
`supersymmetry' transformation the q's in the 
second generation representation couple to a
Tr unit of 4! with no massless generators.

A cancellation occurs in the first generation
component between the q's and the q*'s. The 
coupled script identities in the first generation
are considered equivalent to the coupled I's
and don't cancel.
The scalar Higgs components from the table
for the charm quark are 7.7!+3.6! and
an explicit calculation of the U(1) Higgs
components gives a value of 4.(4!-2).
The calculation gives a mass $D^0$=1864.39MeV.
(empirical value 1864.6(0.5).
A similar calculation of the $D^{\pm}$
mass gives a value of 1867.56MeV
(the empirical value is 1869.3(0.5).

\section{Charmed strange}

Again this meson contains only second generation
quarks at most so $G=2$. The strange quark requires
only one second generation representation and the
charm requires two so $R=3$. The superparity
numbers are the same as for the D mesons;

\vspace{0.5cm}

\centerline
{\epsfig
{file=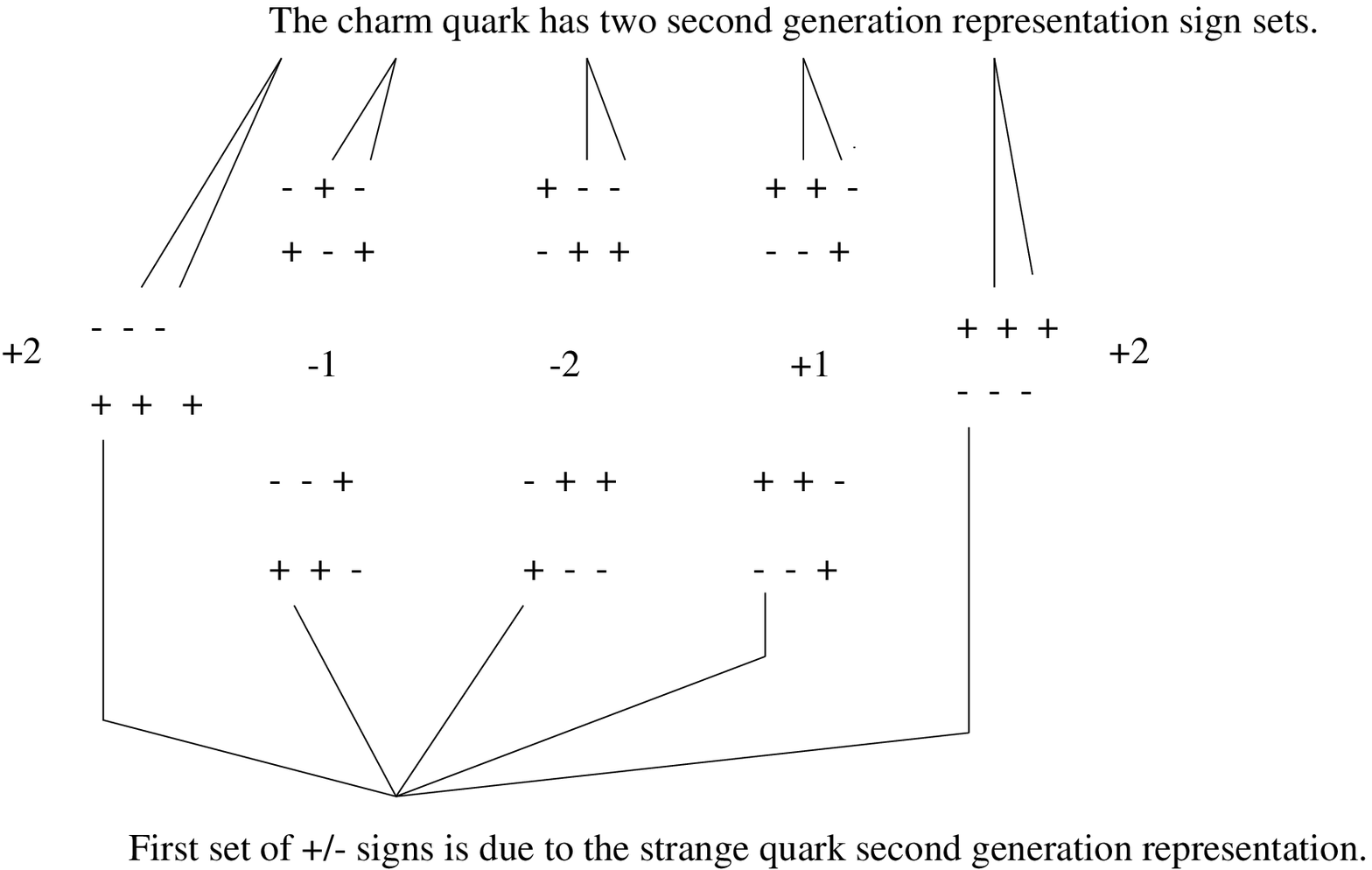,
width=10cm}}

\vspace{0.5cm}

In the above diagram the left hand sign-pair
of each triplet is due to a single component in
the second generation strange representation and
the remaining two sign pairs are due to
the charm. The quantum numbers which the
particle vector must represent can be calculated
from this diagram and are as follows;

\vspace{0.5cm}

\begin{center}
\begin{tabular}{|c|c|c|c|c|c|c|c|c|c|c|c|c|}
\multicolumn{13}{c}{C-S meson Structure;
Particle Vector}
\\\hline
&\multicolumn{2}{c|}{$\mp\;\;\mp\mp$}
&\multicolumn{2}{c|}{$\mp\;\;\pm\mp$}
&\multicolumn{2}{c|}{$\mp\;\;\pm\pm$}
&\multicolumn{2}{c|}{$\pm\;\;\mp\mp$}
&\multicolumn{2}{c|}{$\pm\;\;\pm\mp$}
&\multicolumn{2}{c|}{$\pm\;\;\pm\pm$}
\\$C/{\bar{C}}$ `parity'S 2C 
&\multicolumn{2}{c|}{$\;$}&
\multicolumn{2}{c|}{$\mp\;\;\mp\pm$}
&\multicolumn{2}{c|}{$\;$}&\multicolumn{2}{c|}{$\;$}
&\multicolumn{2}{c|}{$\pm\;\;\;\mp\pm$}
&\multicolumn{2}{c|}{$\;$}
\\\hline
Strangeness&-1&$\;$&-1
&$\;$&-1&$\;$&+1&$\;$&+1
&$\;$&+1&$\;\;$\\\hline
 charm&$\;$&+1&$\;$&-1
&$\;$&+1&$\;$&+1&$\;$&-1&$\;$&+1\\\hline
Superparity&\multicolumn{2}{c|}{+2}
&\multicolumn{2}{c|}{-1}
&\multicolumn{2}{c|}{-2}
&\multicolumn{2}{c|}{-2}
&\multicolumn{2}{c|}{-1}
&\multicolumn{2}{c|}{+2}
\\\hline
$\Sigma$&\multicolumn{2}{c|}{+2}
&\multicolumn{2}{c|}{-3}
&\multicolumn{2}{c|}{-2}
&\multicolumn{2}{c|}{0}
&\multicolumn{2}{c|}{-1}
&\multicolumn{2}{c|}{+4}
\\\hline
\end{tabular}
\end{center}

\vspace{0.5cm}

Points to note;

1;
The strangeness assignment is given by
the first sign pair of any triplet;
$\pm=+1$ and $\mp=-1$.

2;
The charm assignment is a function of the
remaining two sets of sign-pairs;
$\pm\pm$ or $\mp\mp$ is an even permutation
of charm $+1$ and $\mp\pm$ or $\pm\mp$
is an odd permutation and defined as charm
of -1.

3;
The superparity of an odd charm pair (charm=-1)
is zero. The superparity assignment of the 
strange is 1 with the sign  depending on whether it is
an odd or even in sign with respect to the
charm. The superparity assignment of the
even charm permutation pair (charm = +1) is 1.
Thus an even combination (such as $\pm\pm\pm$)
has superparity +2 whilst an odd combination
(such as $\mp\pm\pm$) has superparity -2. The
superparity is found by summing over the two
quarks. Where the absolute value of the 
superparity of the charm is zero (i.e. for
odd charm pair permutations charm = -1) then
the single strange is odd by definition and
has a superparity of -1.

4;
The $\Sigma$ value is the sum of the 
strangeness assignment, charm and superparity
for a given triplet of signs. S is the number of
{\it{different}} values of $\Sigma$ so in this case
S=6.

5;
P is the number of different sign triplets;
here P=8. R is the number of entries in
the sign blocks; here we have triplets
so R=3. Clearly for only second generation
quarks G=2. 

Plugging in the values for the
constituent quark and gluon energy;
\[
G.P.3.(4!-2).4!+R.S.2.(4!)^2=46080
\]
which is the matrix order which may be
interpreted as the number of possible
distinct discrete states that are possible
particle vector representations.

Now we must calculate the current quark mass.

\begin{center}
\begin{tabular}{ccc}
\multicolumn{3}{c}{C-S meson; Current Quarks
(first/second generation).}
\\\hline\\
&$\;\;\;\;
$strong component$\;\;\;\;$&E.M. component\\
\\
red$\rightarrow$&${\cal{I}}\;\;\;\;\;\;\;\;\;\;\;q$
&$I\;\;\;\;\;\;\;\;\;\;\;\;\;q*$\\
green$\rightarrow$&$q*\;\;\;\;\;\;\;\;\;\;I$
&$q*\;\;\;\;\;\;\;\;\;\;\;\;I$\\
blue$\rightarrow$&$q*\;\;\;\;\;\;\;\;\;\;I$
&$q*\;\;\;\;\;\;\;\;\;\;\;\;I$\\\\
&$\;\;\;\;\;{\stackrel{\uparrow}{\mbox{charm}}}\;
\;{\stackrel{\uparrow}
{\mbox{anti-strange}}}$
&$\;\;\;\;\;{\stackrel{\uparrow}{\mbox{charm}}}\;
\;{\stackrel{\uparrow}
{\mbox{anti-strange}}}$\\\\\hline
\end{tabular}
\end{center}

The first and second generation representations are
identical although for the second generation rep.
 it is the identities
in the E.M. matrix which couple to the $\bar{C}$
generators whilst the $q^*$ in the E.M. matrix
are un-coupled from the generators and are
acting as scalars. Unit E.M. charge is defined
by coupling to the G=1 part of the particle vector.
We need, however, to calculate the mass of only
one set of representations of the current quark
operators since only one set is needed to couple
to both the G=1 and G=2 parts of the particle vector.

As usual the masses are read off the chart. 
The triplet of $q^*$'s couples to a $T_r$
equivalent unit of order (4!-2) and all
other components couple to order 4! One $q$
cancels a $q^*$ and one $\cal{I}$ cancels an $I$
in the strong matrix. There is a doubling of
the order to couple to reverse particle vector
formats giving a total matrix order of;
\[
10.4!+6.(4!-2)
\] 
which, along with the 46080 matrices of the 
constituent mass will be given a standard
non-perturbative radiative correction
${\cal{R}}=1+\alpha_{em}+\alpha_{GF}\approx1.0073115.$

To this the appropriate Higgs components for the 
two current quarks must be added. These are read off
the Higgs chart as a matrix equivalent order.
The energy of potential separation of the two quarks
is also read off the standard chart; both the charm and
the strong have values -2 for the second generation
representation. There is no first generation representation
of the current quarks so the appropriate matrix
order is $|-2-2|(4!-2)$ with no radiative correction
($\alpha_s=1$ at the $T_c$ or $T_r$ scale).
Thus the C-S ground state meson has the composition;

\vspace{0.5cm}

\centerline
{\epsfig{file=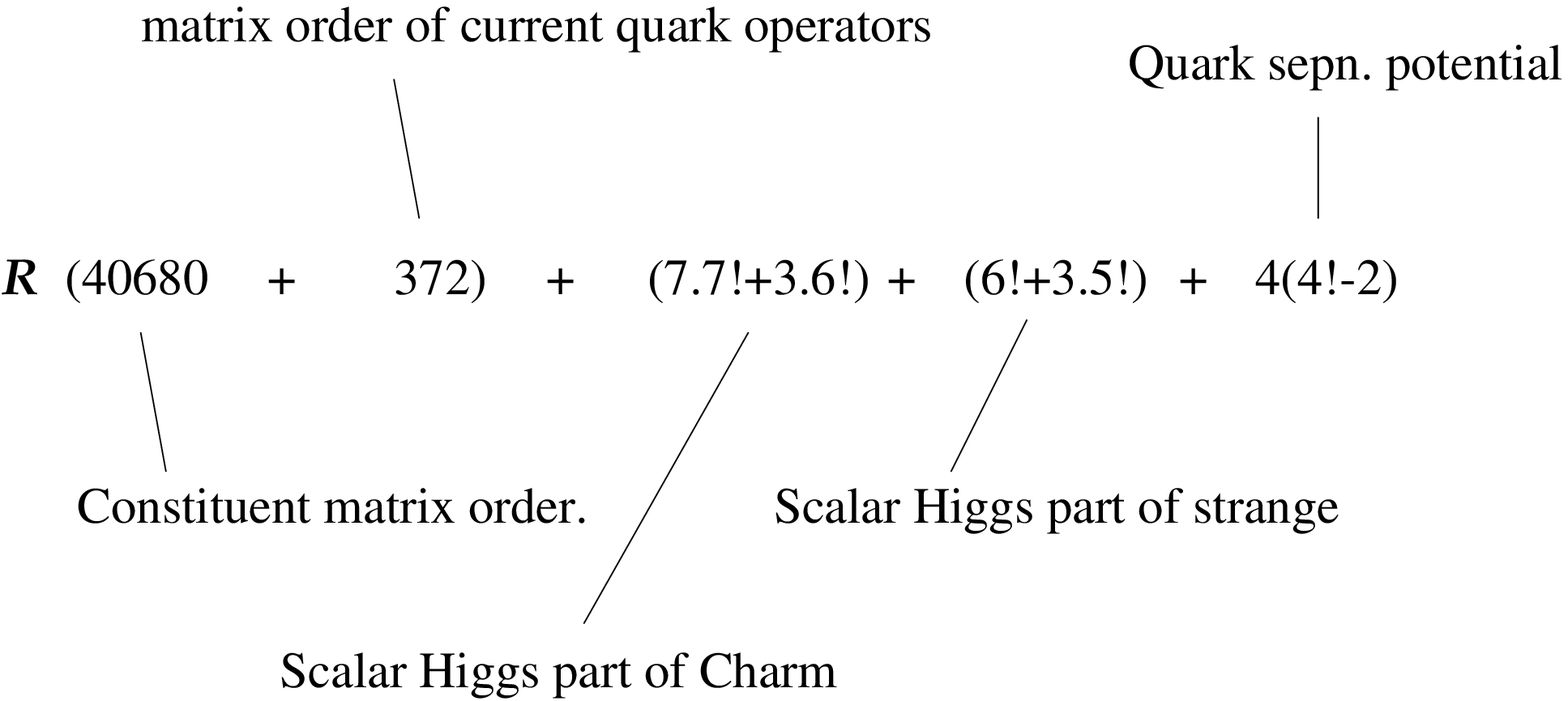,width=10cm}}

\vspace{0.5cm}

Which gives the charmed-strange meson mass
as 1969.12MeV (1968.6(0.6){\cite{groom}}) which is  within
the empirical boundaries.

\section{B mesons}

In this section we will evaluate a B meson 
mass as an example of bottom containing
calculations.

The combination of an up or down
quark with a bottom quark will
require a generation number of G=3
to define a particle vector. To create a
bottom quark we duplicate the
second generation representation and
keep it in phase with the third representation
in terms of parity and superparity 
assignments. The following tabulation
will illustrate the ideas involved;

\begin{center}
\begin{tabular}{|c|c|c|c|c|c|c|c|c|}
\multicolumn{9}{c}{B meson Structure;
Particle Vector}
\\\hline 1st, 2nd \& 3rd gen.
&\multicolumn{2}{c|}{$\mp\;\;\mp\;\;\mp$}
&\multicolumn{2}{c|}{$\mp\;\;\pm\;\;\pm$}
&\multicolumn{2}{c|}{$\pm\;\;\pm\;\;\pm$}
&\multicolumn{2}{c|}{$\pm\;\;\mp\;\;\mp$}
\\$C/{\bar{C}}$ comb. 
&\multicolumn{2}{c|}{$\;$}&\multicolumn{2}{c|}{$\;$}
&\multicolumn{2}{c|}{$\;$}&\multicolumn{2}{c|}{$\;$}
\\\hline
(1st gen.) parity&+1&$\;$&+1
&$\;$&-1&$\;$&-1&$\;$\\\hline
B.ness = $\Sigma$ 2nd \& 3rd.
&$\;$&-2&$\;$&+2
&$\;$&+2&$\;$&-2\\\hline
Superparity&\multicolumn{2}{c|}{-3}
&\multicolumn{2}{c|}{-3}
&\multicolumn{2}{c|}{+3}
&\multicolumn{2}{c|}{+3}
\\\hline
Summed Numbers&\multicolumn{2}{c|}{-4}
&\multicolumn{2}{c|}{0}
&\multicolumn{2}{c|}{+4}
&\multicolumn{2}{c|}{0}
\\\hline
\end{tabular}
\end{center}

The Bottomness here is the sum of the
`parity' of the second and third generation
components and is rather like an extension
of strangeness as it has the same sign 
configuration. Notice that for the 
Bottom quark the second and third generation
components are always `in phase'.

The superparity is related to the total 
number of $\pm$ or $\mp$'s and is
odd or even according to the
product
 e.g.; $\pm\mp\mp$=even=+3. The
summed numbers are self explanatory.

From the above table we can read off 
 P=4 and S=3. R in this case is 2
not 3 because the representations of 
the second and third generation components
are identical in each case.

Thus the particle order is;
\begin{equation}
R.S.2.24^2+G.P.3.22.24=25920
\end{equation}
for two colour-exchange equivalent gluons and three
possible colours of particle vector.

The first-generation current quark mass
chart is identical to that for
the Kaons apart from the scalar pieces.
 This was given as ${\cal{R}}
(10.4!+6.\{4!-2\})+\mbox{scalar part}$
for the $B^{\pm}$.

The second generation and third generation
components of the B meson are identical
duplicates of the second generation
component for the Kaons so the corresponding
mass is  tripled (the third generation 
component is double valued with respect to
the second). Thus the sum of the second
and third generation pieces for the 
charged B contributes ${\cal{R}}(12.4!+12.\{4!-2\})$.

From the tables of Higgs scalar components the
B quark has components 5.8! plus U(1) contributions.
The U(1) contribution is 
assumed to be four times the
up-down current and thus =4(4!-2);
i.e. doubled on the coupling to the third
generation representation. The
charged B mass is then 5276.1MeV
(the empirical value is 5278.9(1.8)MeV).
A similar calculation gives the $B^0$ mass
as 5280.5MeV (5279.2(1.8)MeV empirical). Note
that we seem to have an identical situation to
that with the pions; the up quark appears
to have scalar Higgs pieces single SU(2) and single
U(1) in the charged B meson. If these are included
the predicted charged B mass would increase to 5279.4MeV.

\section{Bottom-strange meson}

As for the case of the charmed-strange
the first generation component of the
current quarks of the strange-bottom
will not acquire mass because  of the absence
of any massive up or down current-quark
content. However the particle vector
continues to reflect the summing over
generation representations and the G number
will be 3. A strange and a bottom can be
represented with R=2. The SQN's are found
from the following chart of quantum
numbers;

\begin{center}
\begin{tabular}{|c|c|c|c|c|c|c|c|c|}
\multicolumn{9}{c}{B-S meson Structure;
Particle Vector}
\\\hline  2nd, 2nd \& 3rd gen.
&\multicolumn{2}{c|}{$\mp\;\;\mp\;\;\mp$}
&\multicolumn{2}{c|}{$\mp\;\;\pm\;\;\pm$}
&\multicolumn{2}{c|}{$\pm\;\;\pm\;\;\pm$}
&\multicolumn{2}{c|}{$\pm\;\;\mp\;\;\mp$}
\\$C/{\bar{C}}$ comb.
&\multicolumn{2}{c|}{$\;$}&\multicolumn{2}{c|}{$\;$}
&\multicolumn{2}{c|}{$\;$}&\multicolumn{2}{c|}{$\;$}
\\\hline
S.ness&-1&$\;$&-1
&$\;$&+1&$\;$&+1&$\;$\\\hline
B.ness = $\Sigma$ 2nd and 3rd 
&$\;$&-2&$\;$&+2
&$\;$&+2&$\;$&-2\\\hline
Superparity&\multicolumn{2}{c|}{-3}
&\multicolumn{2}{c|}{-3}
&\multicolumn{2}{c|}{+3}
&\multicolumn{2}{c|}{+3}
\\\hline
Summed Q Numbers&\multicolumn{2}{c|}{-6}
&\multicolumn{2}{c|}{-2}
&\multicolumn{2}{c|}{+6}
&\multicolumn{2}{c|}{+2}
\\\hline
\end{tabular}
\end{center}

Thus G=3, R=2 (the two bottom sets of
$\pm$ signs - the two on the right
of each box - are always `in phase' and
are treated as a single rep.),
P=4 (the number of different boxes
of sets of $\pm$  signs)
and S=4 because there are 4 different
SQN's.

Hence the constituent particle
vector order is 3.4.3.(4!-2).4!
and the corresponding glue order
is $4.2.2.(4!)^2$.

For the current quarks we note that
the bottom will sum over both 
higher generation reps. whilst the 
strange will only have a single 
higher generation rep. Thus for a
second generation representation of 
the current quark content of the 
neutral S-B
we will have;

\begin{center}
\begin{tabular}{ccc}
\multicolumn{3}{c}{$S{\bar{B}}$ meson; Current Quarks
(first/second generation).}
\\\hline\\
&$\;\;\;\;
$strong component$\;\;\;\;$&E.M. component\\
\\
red$\rightarrow$&${q^*}\;\;\;\;\;\;\;\;\;\;\;q$
&$q\;\;\;\;\;\;\;\;\;\;\;\;\;q*$\\
green$\rightarrow$&$I\;\;\;\;\;\;\;\;\;\;I$
&$I\;\;\;\;\;\;\;\;\;\;\;\;I$\\
blue$\rightarrow$&$I\;\;\;\;\;\;\;\;\;\;I$
&$I\;\;\;\;\;\;\;\;\;\;\;\;I$\\\\
&$\;\;\;\;\;{\stackrel{\uparrow}{\mbox{strange}}}\;
\;{\stackrel{\uparrow}
{\mbox{anti-bottom}}}$
&$\;\;\;\;\;{\stackrel{\uparrow}{\mbox{strange}}}\;
\;{\stackrel{\uparrow}
{\mbox{anti-bottom}}}$\\\\\hline
\end{tabular}
\end{center}

The $q$ cancels the $q^*$ and the mass of
the identities is 8.4! in the first generation
and 4.4!+4.(4!-2) in the second generation
(because of the super-symmetry transform the
identities couple to the spinor generators in
the second generation rep.). These values then double
for the coupling to $C{\bar{C}}$/${\bar{C}}c$ pairs
in the particle vector.

The bottom alone has a third generation rep.;

\begin{center}
\begin{tabular}{ccc}
\multicolumn{3}{c}{$S{\bar{B}}$ meson; Current Quarks
(3rd. generation).}
\\\hline\\
&$\;\;\;\;
$strong component$\;\;\;\;$&E.M. component\\
\\
red$\rightarrow$&$\;\;\;\;\;\;\;\;\;\;\;\;\;q$
&$\;\;\;\;\;\;\;\;\;\;\;\;\;\;\;q*$\\
green$\rightarrow$&$\;\;\;\;\;\;\;\;\;\;\;\;I$
&$\;\;\;\;\;\;\;\;\;\;\;\;\;\;I$\\
blue$\rightarrow$&$\;\;\;\;\;\;\;\;\;\;\;\;I$
&$\;\;\;\;\;\;\;\;\;\;\;\;\;\;I$\\\\
&$\;\;\;\;\;{\stackrel{\uparrow}{\mbox{strange}}}\;
\;{\stackrel{\uparrow}
{\mbox{anti-bottom}}}$
&$\;\;\;\;\;{\stackrel{\uparrow}{\mbox{strange}}}\;
\;{\stackrel{\uparrow}
{\mbox{anti-bottom}}}$\\\\\hline
\end{tabular}
\end{center}

The mass equivalence of the above table is 
of double value again with respect to the 
previous table and is thus 8.4! + 8.(4!-2)
because the $q$ again cancels the $q^*$ and
the identities on the right couple to the
spinor generators.

From the Higgs table the scalar mass of the
bottom is again 5.8! and for the charm 6!+3.5!
whilst the potential energy of separation of
the current quarks (the U(1) terms) are based
on bond between second to second and second 
to third generation reps. The second to second
bond for the BS will be double the down-down.
(The d-d is 4.(4!-2)). The second to third
bond is double again giving net \{16+8\}.(4!-2).

With the radiative correction we have
a total mass matrix order;
\[
{\cal{R}}.{(29344)+{\mbox{current quark mass}}}
\]
Which gives a value 5.367GeV using the given values
for the current quarks  and
the Higgs components for the bottom and the strange 
from the chart. The empirical value is 5.369(24){\cite{groom}}.

\section{Charmed bottom.}

\begin{center}
\begin{tabular}{|c|c|c|c|c|c|c|c|c|c|c|c|c|}
\multicolumn{13}{c}{C-B meson Structure;
Particle Vector}
\\\hline
&\multicolumn{2}{c|}{$\mp\;\;\mp\mp$}
&\multicolumn{2}{c|}{$\mp\;\;\pm\mp$}
&\multicolumn{2}{c|}{$\mp\;\;\pm\pm$}
&\multicolumn{2}{c|}{$\pm\;\;\mp\mp$}
&\multicolumn{2}{c|}{$\pm\;\;\pm\mp$}
&\multicolumn{2}{c|}{$\pm\;\;\pm\pm$}
\\$C/{\bar{C}}$;B,2C 
&\multicolumn{2}{c|}{$\;$}&
\multicolumn{2}{c|}{$\mp\;\;\mp\pm$}
&\multicolumn{2}{c|}{$\;$}&\multicolumn{2}{c|}{$\;$}
&\multicolumn{2}{c|}{$\pm\;\;\;\pm\mp$}
&\multicolumn{2}{c|}{$\;$}
\\\hline
B.ness&-2&$\;$&-2
&$\;$&-2&$\;$&+2&$\;$&+2
&$\;$&+2&$\;\;$\\\hline
 charm&$\;$&+1&$\;$&-1
&$\;$&+1&$\;$&+1&$\;$&-1&$\;$&+1\\\hline
Sp.parity&\multicolumn{2}{c|}{-3}
&\multicolumn{2}{c|}{+2}
&\multicolumn{2}{c|}{+3}
&\multicolumn{2}{c|}{-3}
&\multicolumn{2}{c|}{+2}
&\multicolumn{2}{c|}{+3}
\\\hline
$\Sigma$&\multicolumn{2}{c|}{-4}
&\multicolumn{2}{c|}{-1}
&\multicolumn{2}{c|}{+2}
&\multicolumn{2}{c|}{0}
&\multicolumn{2}{c|}{+3}
&\multicolumn{2}{c|}{+6}
\\\hline
\end{tabular}
\end{center}

Notes;
1; The first of every three sets of $\pm$ signs is
due to a {\it{double valued}} bottom. It represents
the relative orientation of the $C/\bar{C}$ in the
combined second and third generations which are 
always in sync. The other pair of $\pm$ signs is due
to the second generation representation of the charm
quark. 

2; The bottomness quantum number is defined from
the first of each three sets of signs so that
$\mp\;**$ is the particle vector potential 
to express bottomness -2 (this quantum number is 
double valued in discretised QCD as mentioned - it
does NOT mean a bottomness of 2 in conventional
theory when the number is normalised to unity)
and $\pm\;**$ is bottomness +2.

3; as previously the charm quantum number is defined
in terms of odd or even permutations of its' second
generation reps. so that $*\;\pm\pm$ 
or $*\;\mp\mp$ is
+1 unit of charm and $*\;\pm\mp$ or $*\;\mp\pm$ is 
-1 unit of charm.

4; The superparity number of
the negative charm quantum number is 0; that is
the charm product
$\pm\mp$ or $\mp\pm$ has zero superparity. The charm
combination $\pm\pm$ has superparity +1 and
the combination $\mp\mp$ has superparity -1.
The bottomness
quantum number has a superparity $|2|$ value
with the sign following the quantum number bottomness.
The given superparity states are then the corresponding
sums of superparity.

5; The $\Sigma$ is then the sum of the bottomness, charm
and superparity quantum numbers. here S = 6 is the
number of different $\Sigma$ values and represents the 
number of distinct matrix representations seen by the
gluons.

In the case of the ground state C-B meson G = 3
is the generation number. The P number is the 
number of distinct sets of sign triplets and 
here P = 8. R is the representation number and
parallels the quantum numbers R =  1 for the
charm and R = 2 for the bottom in this chart
so that the total is R = 3.

Thus the constituent matrix order (number of 
distinct matrices in the particle vector and
gluon representations);

\[
\mbox{Constituent quarks} = G.P.3.4!.(4!-2) =38016
\]

\[
\mbox{Constituent glue} = R.S.2.(4!)^2=20736\]

The current quark reps. look identical to those of
the Kaon with the C replacing the up and
the B replacing the strange  but remember
that this rep. is duplicated for the second
generation representation with the appropriate
supersymmetry transformation of the operators;

\begin{center}
\begin{tabular}{ccc}
\multicolumn{3}{c}{C-B; Current Quarks
(2nd generation rep.).}
\\\hline\\
&$\;\;\;\;
$strong component$\;\;\;\;$&E.M. component\\
\\
red$\rightarrow$&${\cal{I}}\;\;\;\;\;\;\;\;\;\;\;q$
&$I\;\;\;\;\;\;\;\;\;\;\;\;\;q*$\\
green$\rightarrow$&$q*\;\;\;\;\;\;\;\;\;\;I$
&$q*\;\;\;\;\;\;\;\;\;\;\;\;I$\\
blue$\rightarrow$&$q*\;\;\;\;\;\;\;\;\;\;I$
&$q*\;\;\;\;\;\;\;\;\;\;\;\;I$\\\\
&$\;\;\;\;\;{\stackrel{\uparrow}{\mbox{charm}}}\;
\;{\stackrel{\uparrow}
{\mbox{anti-bottom}}}$
&$\;\;\;\;\;{\stackrel{\uparrow}{\mbox{charm}}}\;
\;{\stackrel{\uparrow}
{\mbox{anti-bottom}}}$\\\\\hline
\end{tabular}
\end{center}

The matrix order here is $2\{10.4!+6.(4!-2)\}$
where the multiplier 2 arises because we are summing
over first and second generation representations.
The bottom will additionally have  a third generation
rep. which will have a mass of $2(4.4!+4\{4!-2\})$

The `static' purely strong energy of separation of 
the quarks using the table;

\vspace{0.5cm}

\begin{center}
\begin{tabular}{|c|c|c|c|c|c|c|c|c|c|c|c|c|}
\multicolumn{13}{c}{chart of static strong potentials}
\\\hline
&\multicolumn{2}{c|}{up}
&\multicolumn{2}{c|}{down}
&\multicolumn{2}{c|}{charm}
&\multicolumn{2}{c|}{strange}
&\multicolumn{2}{c|}{top}
&\multicolumn{2}{c|}{bottom}
\\\hline
First generation
&\multicolumn{2}{c|}{+1}
&\multicolumn{2}{c|}{-2}
&\multicolumn{2}{c|}{+1}
&\multicolumn{2}{c|}{-2}
&\multicolumn{2}{c|}{+2}
&\multicolumn{2}{c|}{-4}
\\\hline
Second generation
&\multicolumn{2}{c|}{0}
&\multicolumn{2}{c|}{0}
&\multicolumn{2}{c|}{-2}
&\multicolumn{2}{c|}{-2}
&\multicolumn{2}{c|}{0}
&\multicolumn{2}{c|}{0}
\\\hline
Third generation
&\multicolumn{2}{c|}{0}
&\multicolumn{2}{c|}{0}
&\multicolumn{2}{c|}{0}
&\multicolumn{2}{c|}{0}
&\multicolumn{2}{c|}{?+4}
&\multicolumn{2}{c|}{?-4}
\\\hline
\end{tabular}
\end{center}

\vspace{0.5cm}

Using the above table we have potential
first generation rep. to first for
the CB as $|+1-4|(4!-2)$ and first to third
an identical value while second to third potential 
$|-2-4|(4!-2)$ with an identical second to
first giving a net $18(4!-2)$. The 
Higgs values from the charts are
\[
\mbox{charm}=6!+3.5!\]
and
\[
\mbox{bottom}=5.8!\]

These last values i.e. the $18.(4!-2)+5.8!+6!+3.5!$
have no radiative corrections. The remaining mass
is given a $T_r$ scale radiative correction
${\cal{R}}=1+\alpha_{em}+G_f$
and the result of the calculation is

Mass C-B =  6.9116GeV

\noindent
which unfortunately is well above the empirical
value of about 6.4GeV{\cite{groom}}.

\section{Top-antidown and Top-antiup meson  ground states.}

We saw in the case of the B mesons that
the second generation component of the
B quark is a `dummy' duplicate; it doesn't
contribute to the R representation number.
Thus we can use the model of the charm-containing
meson to produce a calculation of the T
mesons by extending the the D meson calculations
just as an extension of the K calculations gives
the B meson masses. The top quark mass however
is overwhelmingly dominated by the Higgs scalar
components of the top quark which, from the table, are 
$2.9.9!+10.8!\approx159.9GeV$. 
With discretised Q.C.D. we can do a precision
calculation of the ground state mass containing a 
top quark but the result varies little from the
top mass for combination with light quarks. Such combinations
are unlikely to be stable. Nevertheless the calculation can
be done as an exercise.

Clearly G=3 for a top/anti-up or top/anti-down
combination. To get the R,P and S values we
need a table of possible quantum states
(only the first entry of the $C/{\bar{C}}$ is given;
the reader can fill in the remaining values if 
interested!);

\begin{center}
\begin{tabular}{|c|c|c|c|c|c|c|c|c|c|c|c|c|}
\multicolumn{13}{c}{$T\bar{U}$ , $T\bar{D}$ mesons Structure;
Particle Vector}
\\\hline
&\multicolumn{2}{c|}{$\mp\;\;\;\;\mp\mp\;\mp\mp$}
&\multicolumn{2}{c|}{$$}
&\multicolumn{2}{c|}{$$}
&\multicolumn{2}{c|}{$$}
&\multicolumn{2}{c|}{$$}
&\multicolumn{2}{c|}{$$}
\\$C/{\bar{C}}$
&\multicolumn{2}{c|}{$\;$}&
\multicolumn{2}{c|}{$$}
&\multicolumn{2}{c|}{$\;$}&\multicolumn{2}{c|}{$\;$}
&\multicolumn{2}{c|}{$$}
&\multicolumn{2}{c|}{$\;$}
\\\hline
(1st Gen)&+1&$\;$&+1
&$\;$&+1&$\;$&-1&$\;$&-1
&$\;$&-1&$\;\;$\\\hline
Top&$\;$&+2&$\;$&-2
&$\;$&+2&$\;$&+2&$\;$&-2&$\;$&+2\\\hline
Sup.p'&\multicolumn{2}{c|}{-3}
&\multicolumn{2}{c|}{-1}
&\multicolumn{2}{c|}{-3}
&\multicolumn{2}{c|}{+3}
&\multicolumn{2}{c|}{+1}
&\multicolumn{2}{c|}{+3}
\\\hline
$\Sigma$&\multicolumn{2}{c|}{0}
&\multicolumn{2}{c|}{-2}
&\multicolumn{2}{c|}{0}
&\multicolumn{2}{c|}{+4}
&\multicolumn{2}{c|}{-2}
&\multicolumn{2}{c|}{+4}
\\\hline
\end{tabular}
\end{center}

Because of the dummy duplication of the second generation
representation we take R=3. Obviously P=8 and S=3.
Thus the particle and glue order is
48384 (to be multiplied by $\cal{R}$).
The topness number is defined in much the
same way as charm but this in not the top quantum
number as such; this is carried in the current
quark representation. (The particle vector sums
over possible states and that is why it includes
top quantum numbers with opposite signs; just
as for bottomness, charm and stangeness in 
the other particle vectors studied).
 We can use the D meson current
quark mass calculations to determine
the current quark mass by doubling the value
of the second generation component for the
D to get the top containing masses.
The charged $T\bar{D}$ mass is then $\approx
161.038$GeV and $T\bar{U}\approx161.037$GeV.

We will leave the calculation of other
top containing meson masses to the reader
as an exercise.

\section{Conclusion}

We have seen that the discretised system, as outlined in 
previous papers and extended here, is capable of 
providing a good method of representing particle masses
and defining symmetries involved in the observed spectrum of
masses. The model has all the overall symmetry features of the
standard model and is predictive. It is interesting that the 
predicted mass of the charmed bottom meson varies significantly
from the early measurements although it is still within two
standard deviations and it will be interesting to see
if the empirical value moves toward the predicted mass as
measurements of the rest mass of the C-B ground state 
improve. 

The weakest area of the theory is the Higgs sector but
this is also the most interesting as it seems to provide the
first empirical evidence that $SU(5)$ is a real symmetry of
nature; albeit of a most unusual kind!

There is quite a lot of scope for further research in this
topic with numerous unsolved problems and a number of
hints in the correspondence with empirical data of new
physics beyond the standard model. In particular it would
be interestion to address the question of the relationship
between the mass calculation, and the mixing patterns present,
and the CKM matrix entries.

%\end{center}
\end{document}